\documentclass{aa}
\usepackage{txfonts}

\usepackage{color}

\usepackage{epstopdf} 

\usepackage{rotating}
\usepackage{amsfonts}

\usepackage{pdflscape}
\usepackage{ulem}

\usepackage{hyperref}

\usepackage{natbib}
\bibpunct{(}{)}{;}{a}{}{,} 

\begin{document}

\title{Long-term magnetic activity of a sample of M-dwarf stars from the HARPS program\thanks{Based on observations made with the HARPS instrument on the ESO 3.6-m telescope at La Silla Observatory under programme ID 072.C-0488(E)}}
\subtitle{I. Comparison of activity indices}

\author{J. Gomes da Silva\inst{1,2}
	\and N.C. Santos\inst{1,2}
	\and X. Bonfils\inst{3}
	\and X. Delfosse\inst{3}
	\and T. Forveille\inst{3}
	\and S. Udry\inst{4}
	}

\institute{Centro de Astrof\'isica, Universidade do Porto, Rua das Estrelas, 4150-762 Porto, Portugal \\ \email{Joao.Silva@astro.up.pt}
	\and Departamento de F\'isica e Astronomia, Faculdade de Ci\^encias da Universidade do Porto, Portugal
	\and UJF-Grenoble 1 / CNRS-INSU, Institut de Plan\'etologie et d'Astrophysique de Grenoble (IPAG) UMR 5274, Grenoble, F-38041, France
	\and Observatoire de Gen\`eve, 51 ch. des Maillettes, 1290 Sauverny, Switzerland
	}

\abstract 	{The search for extra-solar planets similar to Earth is becoming a reality, but as the level of the measured radial-velocity reaches the sub-m\,s$^{-1}$, stellar intrinsic sources of noise capable of hiding the signal of these planets from scrutiny become more important.}
		{Other stars are known to have magnetic cycles similar to that of the Sun. The relationship between these activity variations and the observed radial-velocity is still not satisfactorily understood. Following our previous work, which studied the correlation between activity cycles and long-term velocity variations for K dwarfs, we now expand it to the lower end of the main sequence. In this first paper our aim is to assess the long-term activity variations in the low end of the main sequence, having in mind a planetary search perspective.}
		{We used a sample of 30 M0--M5.5 stars from the HARPS M-dwarf planet search program with a median timespan of observations of 5.2 years. We computed chromospheric activity indicators based on the \ion{Ca}{ii} H and K, H$\alpha$, \ion{He}{i} D3, and \ion{Na}{i} D1 and D2 lines. All data were binned in to average out undesired effects such as rotationally modulated atmospheric inhomogeneities. We searched for long-term variability of each index and determined the correlations between them.}
		{While the $S_{\mathrm{Ca\,II}}$, H$\alpha$, and \ion{Na}{i} indices showed significant variability for a fraction of our stellar sample (39\%, 33\%, and 37\%, respectively), only 10\% of our stars presented significant variability in the \ion{He}{i} index. We therefore conclude that this index is a poor activity indicator at least for this type of stars.
		Although the H$\alpha$ shows good correlation with $S_{\mathrm{Ca\,II}}$ for the most active stars, the correlation is lost when the activity level decreases. This result appears to indicate that the \ion{Ca}{ii}$-$H$\alpha$ correlation is dependent on the activity level of the star.
		The \ion{Na}{i} lines correlate very well with the $S_{\mathrm{Ca\,II}}$ index for the stars with low activity levels we used, and are thus a good chromospheric activity proxy for early-M dwarfs. We therefore strongly recommend the use of the $\ion{Na}{i}$ activity index because the signal-to-noise ratio in the sodium lines spectral region is always higher than for the calcium lines.}
		{}


\maketitle

\section{Introduction}

The increase in precision of the radial-velocity instruments  is leading to the detection of smaller reflex semi-amplitude signals induced by extra-solar planets (e.g. the 1.9 M$_\oplus$ planet discovered by \citealp{mayor2009}). Also, as the observation timespan increases, the detection of planets in longer orbital periods is becoming possible \citep[e.g.][]{wright2008}.

As these keplerian semi-amplitudes reach the sub-m\,s$^{-1}$ level, the signals induced by intrinsic sources (e.g. oscillations, rotating spots and plages, inhibition of convection) become more significant \citep[e.g.][]{saar1997a,santos2000,paulson2002,bouchy2004,meunier2010,dumusque2011a,dumusque2011b,boisse2011}. It is well-known that stellar activity can be a source of radial-velocity "noise" and can even induce periodic RV signals similar to those produced by extra-solar planets \citep{queloz2001,bonfils2007,huelamo2008}. It is therefore extremely important to study and try to understand how stellar activity produce these signals and how they can be corrected for.

Studies conducted at the Mount Wilson observatory uncovered that many solar-like stars have magnetic cycles similar to that of the Sun \citep{wilson1978,baliunas1995}. These studies concluded that around 85\% of the stars have stellar activity variability \citep{baliunas1998}. Although the Mt. Wilson survey covered a broad range of spectral types, only one M-dwarf was included in the sample. This star, HD\,95735, has an average $S_{MW}$ value of 0.424 and a variable activity cycle (with the cycle period poorly defined as yet). More recently, evidence for cyclic activity was found for a few M dwarfs. \citet{cincunegui2007a} claimed a magnetic cycle with a $\sim$442 day period for the dMe 5.5 Prox Centauri and \citet{diaz2007b} found a P = $\sim$763 day activity period for the dMe 3.5 spectroscopic binary Gl375. Aditionally, activity cycles were also claimed to be present on Gl229A (M1/2) and Gl752A (M2.5) with periods of $\sim$4 yr and $\sim$7 yr, respectively \citep{buccino2010}.
Apart from these few cases, not much is known about the magnetic cycles of the largest population of stars in the solar neighborhood.

In a recent paper we studied a sample of seven late-G and early-K dwarfs with well-known magnetic cycles and compared the activity level of these stars with their radial-velocity variations over a timespan of five years \citep{santos2010}. We found that generally the $S$ index was correlated with H$\alpha$ and anti-correlated with the \ion{He}{i} index. Also, the long-term variations of the $S$ index could be detected in the cross-correlation function parameter's line bisector span, full-width-at-half-maximum and contrast, implying that these parameters could be used to follow the activity cycles of these stars.

In this first paper we extend this study to the lower end of the main sequence and use a more extended sample. We study for the first time the effect of long-term chromospheric activity of a sample of dwarfs of spectral types from M0 to M5.5, covering the range of stars from partially convective to the beginning of fully convective interiors. We used the same HARPS M-dwarf sample as in the search of extra-solar planets in the southern hemisphere \citep[see][submitted]{bonfils2011}.

The outline of this paper is as follows: we first describe our sample and observations in Section \ref{obs}, in Section \ref{indices} we explain the computation of our four activity indices, we search for activity variability in Section \ref{var} and for maxima or minima of activity in Section \ref{trends}, compare the indices in Section \ref{corr}, and our conclusions are finally presented in Section \ref{conclusion}.

\begin{center}
\begin{table*}[htbp]
\caption{Basic parameters of the stellar sample and observation log.}
\label{obslog}
\centering
\begin{tabular}{l l c c c c c c c} \\
\hline
\hline
\multicolumn{1}{l}{Star} &
\multicolumn{1}{l}{Sp. Type\tablefootmark{a}} &
\multicolumn{1}{c}{$V$\tablefootmark{e}} &
\multicolumn{1}{c}{$V-I$\tablefootmark{d}} &
\multicolumn{1}{c}{BJD$_{start}-2400000$ } &
\multicolumn{1}{c}{BJD$_{end}-2400000$} &
\multicolumn{1}{c}{$N_{nights}$}   &
\multicolumn{1}{c}{$T_{span}$}   &
\multicolumn{1}{c}{$<S/N>$}   \\
\multicolumn{1}{c}{} &
\multicolumn{1}{c}{} &
\multicolumn{1}{c}{[mag]} &
\multicolumn{1}{c}{[mag]} &
\multicolumn{1}{c}{[days]} &
\multicolumn{1}{c}{[days]} &
\multicolumn{1}{c}{} &
\multicolumn{1}{c}{[years]} &
\multicolumn{1}{c}{(sp. order 6)} \\
\hline
GJ361	&	M1.5\,V\tablefootmark{b}	&	10.40	&	2.01					&	54455.84 &	55169.84	&	34	&	2.0	&	3.9 \\
GJ2049	&	M1\,V\tablefootmark{e}	&	11.17\tablefootmark{d}	&	1.77 		&	54455.70	&	55155.86	&	25	&	1.9	&	3.2 \\	
GJ3218	&	M2\,V\tablefootmark{e}	&	11.12	&	1.65\tablefootmark{e}	&	54396.89	&	55141.67	&	40	&	2.0	&	2.0 \\
Gl1		&	M3\,V				&	8.57		&	2.13					&	52985.60	&	55048.83	&	43	&	5.6	&	11.5 \\
Gl176	&	M2.5\,V				&	9.97		&	2.25					&	52986.71	&	55132.82	&	70	&	5.9	&	4.2 \\
Gl205	&	M1.5\,V				&	7.92		&	2.08					&	52986.73	&	54386.81	&	75	&	3.8	&	15.0 \\
Gl273	&	M3.5\,V				&	9.89		&	2.71					&	52986.77	&	55126.87	&	61	&	5.9	&	6.5 \\
Gl382	&	M2\,V				&	9.26		&	2.18					&	52986.84	&	54174.65	&	30	&	3.3	&	8.0 \\
Gl393	&	M2\,V				&	9.76\tablefootmark{d}	&	2.26		&	52986.86	&	54883.74	&	28	&	5.2	&	6.9 \\
Gl433	&	M2\,V				&	9.79		&	2.15					&	52989.84	&	55057.47	&	60	&	5.7	&	5.3 \\
Gl436	&	M2.5\,V\tablefootmark{b}	&	10.68	&	2.02					&	53760.83	&	54999.46	&	105	&	3.4	&	2.6 \\
Gl479	&	M3\,V				&	10.64	&	1.90					&	53158.55	&	54571.70	&	57	&	3.9	&	4.2 \\
Gl526	&	M1.5\,V				&	8.46		&	2.07					&	53158.60	&	55001.58	&	34	&	5.0	&	11.6 \\
Gl551	&	M5.5\,V				&	11.05	&	3.62					&	52684.86	&	55057.54	&	37	&	6.5	&	1.2 \\
Gl581	&	M2.5\,V				&	10.57	&	2.53					&	53152.71	&	55056.53	&	128	&	5.2	&	3.6 \\
Gl588	&	M2.5\,V				&	9.31		&	2.40					&	53152.75	&	54956.80	&	25	&	4.9	&	9.2 \\
Gl667C	&	M2\,V				&	10.22	&	2.08\tablefootmark{e}	&	53158.76	&	55053.69	&	147	&	5.2	&	4.0 \\
Gl674	&	M3\,V				&	9.36		&	2.40					&	53158.75	&	54732.48	&	44	&	4.3	&	8.4 \\
Gl680	&	M1.5\,V				&	10.14	&	2.27					&	53159.71	&	55057.65	&	28	&	5.2	&	4.3 \\
Gl699	&	M4\,V				&	9.54		&	2.52					&	54194.89	&	55054.65	&	25	&	2.4	&	6.6 \\
Gl832	&	M1\,V				&	8.67		&	2.18					&	52985.52	&	55122.65	&	57	&	5.9	&	10.4 \\
Gl849	&	M3\,V				&	10.42	&	2.50					&	52990.54	&	55122.62	&	48	&	5.8	&	4.5 \\
Gl876	&	M3.5\,V				&	10.17	&	2.77					&	52987.58	&	55122.63	&	37	&	5.8	&	4.8 \\
Gl877	&	M2.5\,V\tablefootmark{e}	&	10.55\tablefootmark{d}	&	2.43		&	52857.83	&	54704.73	&	39	&	5.1	&	2.8 \\
Gl887	&	M2\,V				&	7.34		&	2.02					&	52985.57	&	54392.68	&	63	&	3.9	&	15.7 \\
Gl908	&	M1\,V				&	8.98		&	2.04					&	52986.58	&	55057.85	&	50	&	5.7	&	11.6 \\
HIP12961	&	M0\,V\tablefootmark{c}	&	9.7		&	1.64					&	52991.63	&	55109.79	&	45	&	5.8	&	3.4 \\
HIP19394	&	M3.5\,V\tablefootmark{e}	&	11.81	&	2.5					&	52942.80	&	55105.79	&	29	&	5.9	&	1.2 \\
HIP38594	&	M\tablefootmark{e}		&	9.96		&	1.64					&	52989.79	&	55105.90	&	16	&	5.8	&	2.8 \\
HIP85647	&	M0\,V\tablefootmark{b}	&	9.59		&	1.85					&	53917.75	&	55117.49	&	27	&	3.3	&	3.8 \\
\hline
\end{tabular}
\tablebib{
(a) \citet[submitted]{bonfils2011} unless individually specified; (b) \citet{hawley1996}; (c) \citet{koen2010}; (d) \citet{esa1997}; (e) Simbad (http://simbad.u-strasbg.fr/simbad/)
}
\end{table*}
\end{center}

\section{Sample and observations}\label{obs}

The sample comes from the HARPS \citep{mayor2003} M-dwarf planet search program, which corresponds to a volume limited selection of stars brighter than V = 14 mag and with a projected rotational velocity $v\sin i \lesssim 6.5$ km\,s$^{-1}$ \citep[see][submitted]{bonfils2011}. They obtained high-resolution spectra (spectral resolution = 115 000) for these stars from 2003 to 2009, with 15 min integration times.

For the calculation of our $S$ index\footnote{From now on the index based on the $\ion{Ca}{ii}$ lines will be referenced as $S$ index or $S_{\mathrm{Ca\,II}}$.} we selected all spectra with a signal-to-noise ratio at spectral order 6 (corresponding to the $\ion{Ca}{ii}$ K line, $\sim$3933 \AA) higher than 2, the value at which the relation between the index and S/N disappears. All other indices were not affected by this selection except for cases where we compared them with $S_{\mathrm{Ca\,II}}$.
The data were then nightly averaged and bins of 150 days were used to suppress any rotationally modulated signals caused by short-term activity. Each bin included at least three nights, and cases where this condition was not met were discarded. At this stage only stars with four or more bins were considered for the rest of this study. This selection enabled us to compare the different parameters and look for correlations with enough data points. From now on, we refer the binned data unless specified. We ended up with a sample of 30 stars matching these conditions (23 for the $S$ index) with spectral types ranging from M0\,V to M5.5\,V.

The errors used for each bin are the errors on the average, $\sigma /\sqrt{N}$, where $\sigma$ is the root-mean-square (rms) of the nightly averaged data in each bin and $N$ the number of measurements used to calculate the average.

Table \ref{obslog} presents some basic parameters and the observation log for these stars.
The start and end dates, number of nights, timespan of observations, and signal-to-noise ratios were calculated for the nightly averaged data without using the $S/N \geq 2$ selection.
The timespan of observations ranges from 1.9 to 6.5 years, which should be sufficient to detect activity cycle variations in these stars if they are present.

\section{Activity indices}\label{indices}

\begin{figure}[htbp]
\begin{center}
\resizebox{\hsize}{!}{\includegraphics{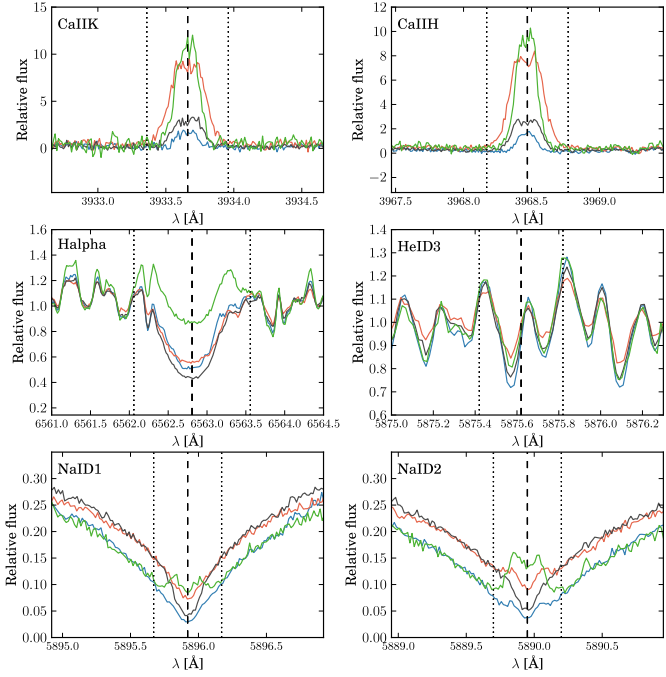}}
\caption{Cores of the lines used to calculate the indices for three stars with similar color ($V-I \sim 2.10$) and different activity levels plus one of the most active stars in the sample (green, Gl479). In blue is Gl1, black is Gl526, and red is Gl205.}
\label{lines}
\end{center}
\end{figure}

The \ion{Ca}{ii} H and K lines are well-known and widely used stellar magnetic activity proxies. Our \ion{Ca}{ii} index was computed as in \citet{boisse2009} and \citet{santos2000} with slight modifications to the wavelength bands at the core of the lines. We chose bands of 0.6 \AA~centered at the \ion{Ca}{ii} H ($\lambda$3968.47) and K ($\lambda$3933.66) lines instead of the 1.09 \AA~band used by \citet{boisse2009} because this last one includes parts of the line wings and thus undesirable photospheric flux (see Figure \ref{lines}, a similar smaller band for the K line was also used by \citet{livingston2007} for the same reasons). The sum of the flux in the two H and K lines was then weighted by the square of their respective errors taken as $\sqrt{N}$ where $N$ is the number of counts inside each band. Our reference bands were two 20 \AA~windows centered at 3900 and 4000 \AA. Because the Mt. Wilson survey only includes one M dwarf, we did not have enough stars to calibrate our $S_{\mathrm{Ca\,II}}$ index to the $S_{MW}$ scale. Furthermore, we could not compute a photospheric free $R'_{HK}$ index \citep{noyes1984} because there is no calibration for stars with $B-V$ higher than 1.2. A comparison of the indices with color is discussed in Section \ref{ind_col}.

The H$\alpha$ index is also a widely used chromospheric indicator, and some authors recommend it for the study of later-type stars owing to the increasing flux in the red part of the spectra for these stars \citep[e.g.][]{pasquini1991,cincunegui2007b}.
Our H$\alpha$ index was also computed as in \citet{boisse2009} but using a slightly broader central band. Instead of the 0.6 \AA~window centered at 6562.808 \AA,~we used a 1.6 \AA~band that seems to include more contribution from chromospheric activity (see Figure \ref{lines}). \citet{pasquini1991} obtained bands in the region 1.6-1.7 \AA~for the chromospheric contribution of the H$\alpha$ line by subtracting the spectra of inactive stars to those of active stars of the same spectral type \citep[the same band was also found by][]{herbig1985}. Recently, \citet{cincunegui2007b} also used a window of 1.5 \AA~to calculate the flux in the H$\alpha$ line. We divided the flux in the central line by the flux in two reference lines of 10.75 and 8.75 \AA~centered at 6550.87 and 6580.31 \AA, respectively.

\citet{diaz2007b} proposed that the \ion{Na}{i} D1 and D2 lines could also be used to follow the chromospheric activity level of very active late-type stars with $(B-V) > 1.1$. In this sample we do not have very active stars, but we decided to study these lines as a complement to the other widely used activity proxies. The \ion{Na}{i} D lines provide information about the conditions in the middle-to-lower chromosphere \citep{mauas2000} and thus are a good complement to the upper chromosphere indicator H$\alpha$ and lower chromosphere proxy \ion{Ca}{ii} index. We first computed our \ion{Na}{i} index in a similar fashion as \citet{diaz2007b} by calculating the average flux in the core of the D1 ($\lambda$5895.92) and D2 ($\lambda$5889.95) lines using 1 \AA~bands.
We used two reference bands with windows of 10 and 20 \AA~centered at 5805.0 and 6090.0 \AA, respectively. For each band we selected the 10 higher flux values and calculated the average. The average values of the two D1 and D2 lines were then divided by the average of the two reference bands.
We noticed that the 1 \AA~bands include a significant part of the line wings and thus possible photospheric contribution to the flux. Therefore, we tried narrower bands of 0.5 \AA~for the lines and noticed that the correlation with the $S_{\mathrm{Ca\,II}}$ index improved drastically (see the results section). These narrower bands were formally adopted for the rest of this work .

The \ion{He}{i} D3 line ($\lambda$5875.62) was observed to show good spatial correlation with plages \citep{landman1981} and can therefore be used as an activity proxy \citep[see][and references therein]{saar1997b,montes1997}.
We used a 0.4 \AA~band centered at the line and divided its flux by the flux in two reference windows of 5 \AA~centered at 5869.0 and 5881.0 \AA,~following the procedure in \citet{boisse2009}.

The errors of the four indices were estimated by differentiating the respective equations, and taking into account the flux in each band.
But because we are interested only on the long-term variations of the indices and we binned the data into 150 day bins, the final errors are the statistical errors on the mean as explained in Section \ref{obs}. The individual errors of the nightly observations are always lower than the estimated statistical errors on the mean used for the binned data.

Figure \ref{lines} shows spectra in the region around the lines used to calculate the activity proxies. The four lines in each plot represent three stars with different activity levels but similar color ($V-I \sim 2.10$) plus one of the most active stars in the sample. By increasing activity, the blue line stands for Gl1 ($<S_{\mathrm{Ca\,II}}>$~$= 0.0147$, $V-I = 2.13$), the black line for Gl526 ($<S_{\mathrm{Ca\,II}}>$~$= 0.0307$, $V-I = 2.07$), the red line represents Gl205 ($<S_{\mathrm{Ca\,II}}>$~$= 0.0920$, $V-I = 2.08$), and the green line illustrates the lines for one of the most active stars in the sample, Gl479 with $<S_{\mathrm{Ca\,II}}>$~$= 0.0960$ but slightly different color ($V-I = 1.90$). The dashed lines mark the line centers, while the dotted lines delimitate the band window. The flux in all figures was normalized by dividing it by the mean flux in the reference bands of each index.

For these four stars the \ion{Ca}{ii} and \ion{Na}{i} lines show a similar behavior. Both increase in emission with increasing activity level in the $S$-index. On the other hand, the H$\alpha$ and \ion{He}{i} lines do not show the same trend. The H$\alpha$ first decreases its emission as activity increases and then exceeds the level of the star with less activity, ultimately arriving at a level similar to that of the continuum (if the activity increases even more it is expected to become an emission line). This behavior was explained by \citet{cram1979} through non-LTE model chromospheres for dK and dM stars. These authors explained that the stronger absorption in the line with increasing chromosphere heating rate is caused by an increase of the n = 2 bound level population of hydrogen and that when the heating reaches a certain level, the chromosphere electron density becomes so high that the line becomes collisionally excited, which increases the emission at its core.
The \ion{He}{i} D3 triplet is observed in Figure \ref{lines} in absorption. This is generally the case for surges, eruptive prominences, and weaker flares, while emission is correlated with more intense flares \citep{zirin1988}.
The behavior observed in the figure is that of an anti-correlation with \ion{Ca}{ii}: as the \ion{Ca}{ii} lines emission increases, there is a tendency for the \ion{He}{i} lines to increase in absorption.

\section{Activity indices variability}\label{var}
After obtaining the four activity indices for our sample we now try to infer which stars show long-term variability and if magnetic activity cycles are present in the data. With this in mind we calculated the rms ($\sigma_e$) and average errors ($<\sigma_i>$) for the indices and compared them using an F-statistics with the F-value computed as $F = \sigma_e^2/<\sigma_i>^2$ following a similar procedure as in \citet{zechmeister2009}.
This approach will give the probability, given an F-value, that the observed variability ($\sigma_e$) can be explained by the mean errors of the data ($\sigma_i$). In this case, the error of the data depends on the high-frequency activity variations of the nightly averaged data and number of bins (as explained in Section \ref{obs}). Therefore, lower values of probability ($P(F)$) will indicate that the variability is not due to the scatter induced by the errors.
The results of this study are presented in Table \ref{table:var}. For the $S$ index we used the $S/N \geq 2$ selection, resulting in $N_{bins}(S)$ bins. Where there are less than four bins, the columns relating to the variability of the $S_{\mathrm{Ca\,II}}$ index appear empty. For the other indices we used no selection, resulting in $N_{bins}$ shown in column 3. The probability that the variation is caused by the errors is presented in the form of a P-value, where values lower than 0.05 (95\% significance) are highlighted in bold.

Because we averaged the data by 150 day bins, we intentionally removed the short-timescale variations, and only long-term variations are to be detected. As a result of this, we expect to find variability in stars with longer periods of observation, where their activity cycle pattern becomes detectable. Indeed, of the four stars with a timespan shorter than three years only one (Gl699) shows significant variability, and just in one of the four indices. Furthermore, from the 18 stars that show variability, 13 have timespans longer than or equal to five years. Indeed, 72\% of the stars with timespans equal to or longer than five years present significant variability in at least one activity index. This is expected because the ability to detect magnetic cycle activity variations should increase with timespan.

Figures \ref{ind_bjd1}, \ref{ind_bjd2}, \ref{ind_bjd3}, \ref{ind_bjd4}, \ref{ind_bjd5}, and \ref{ind_bjd6} show the time-series of the four activity indicators for the selected stars as having significant variability. Small points without errorbars are the data averaged per night, points with error bars represent the binned data, where the errors are the standard error on the mean, the numbers are number of nights used in each bin, and the dashed lines are the second-order polynomials best-fitted to the data. In each plot there is a reference to the peak-to-peak variation ($\Delta$) and to the standard deviations of the binned data ($\sigma$). The values in parenthesis are the percentages of the variations relative to the average values.

Although we may not have enough data to detect full cycles, there are some hints for the cycle phase at which some of these stars could be. For example, in Figure \ref{ind_bjd2} the time-series of $S_{\mathrm{Ca\,II}}$ for Gl433 seems to show a maximum of activity for this star, while for Gl581 a minimum is clearly visible.

\begin{figure*}[!p]
\begin{center}
\resizebox{\hsize}{!}{\includegraphics{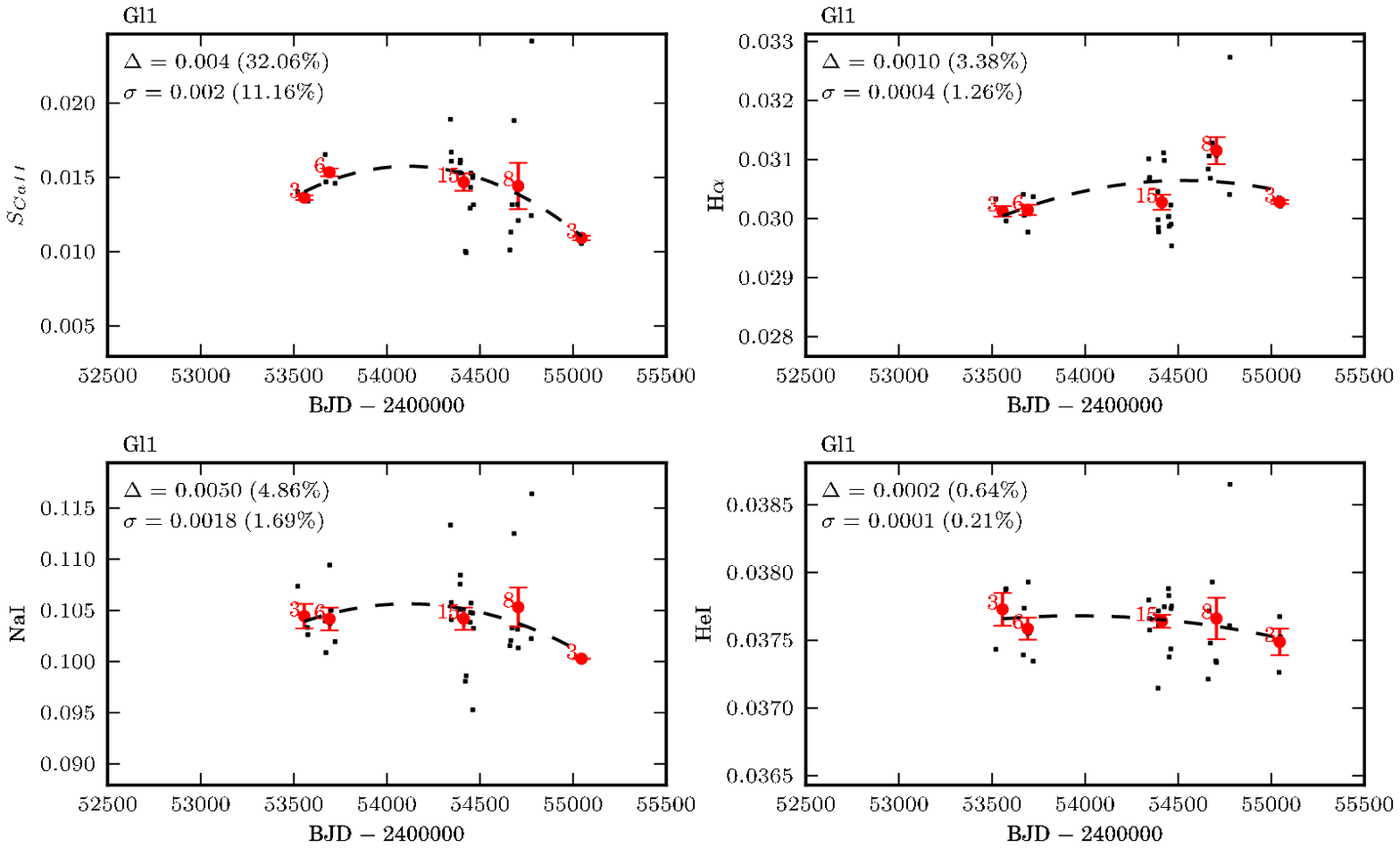}}
\resizebox{\hsize}{!}{\includegraphics{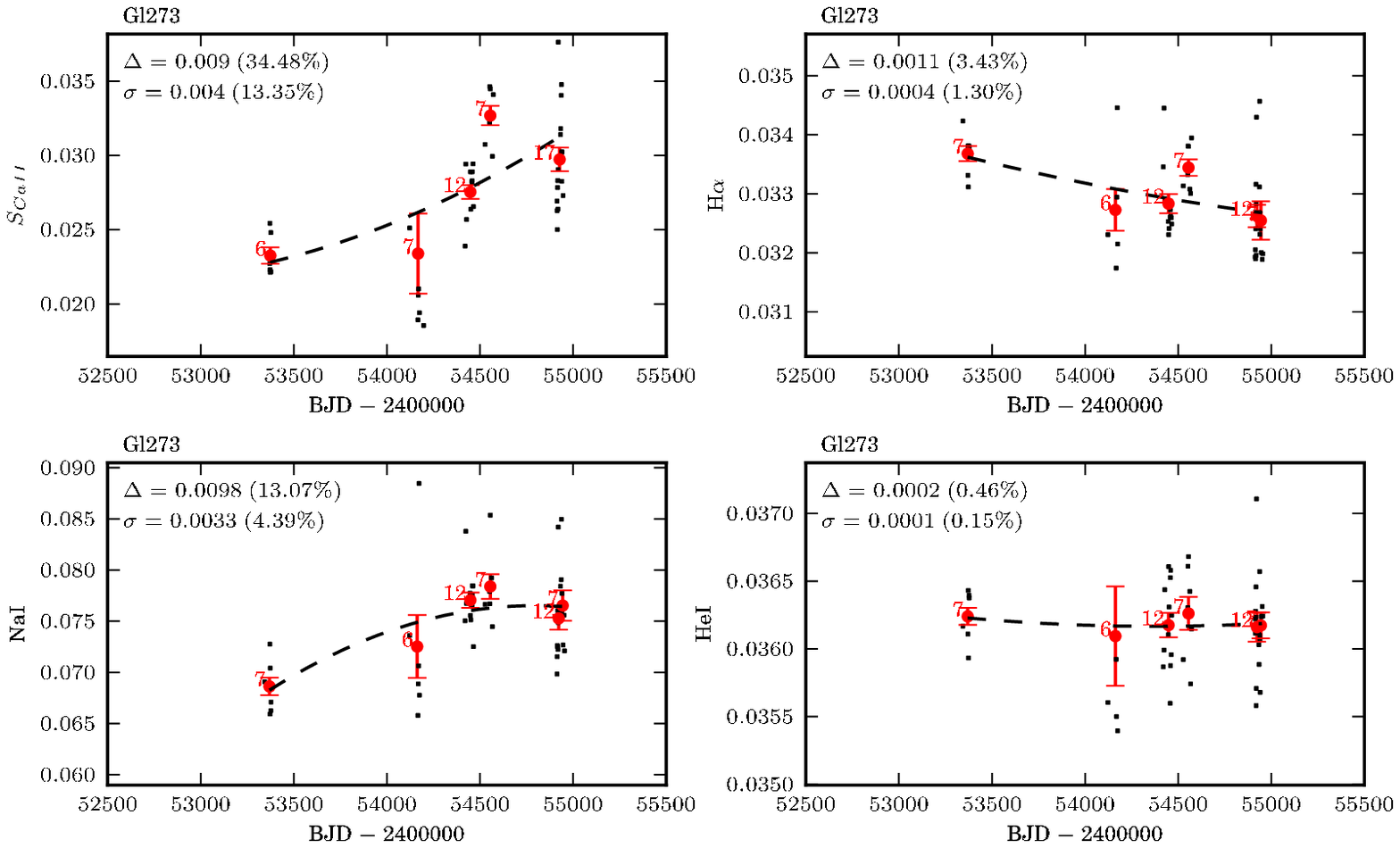}}
\caption{Variations of $S_{\mathrm{Ca\,II}}$, $\ion{H\alpha}{}$, $\ion{Na}{i}$ and $\ion{He}{i}$ with time for the 12 stars with long-term variability (see Section \ref{var}). For visualization purposes, the y-axis is defined as the mean of the data plus and minus $0.35$ times the maximum of the average errors of each index. The x-axis is constant for all stars. Note that Gl877 does not have data for the $S_{\mathrm{Ca\,II}}$ index.}
\label{ind_bjd1}
\end{center}
\end{figure*}

\begin{figure*}[p]
\begin{center}
\resizebox{\hsize}{!}{\includegraphics{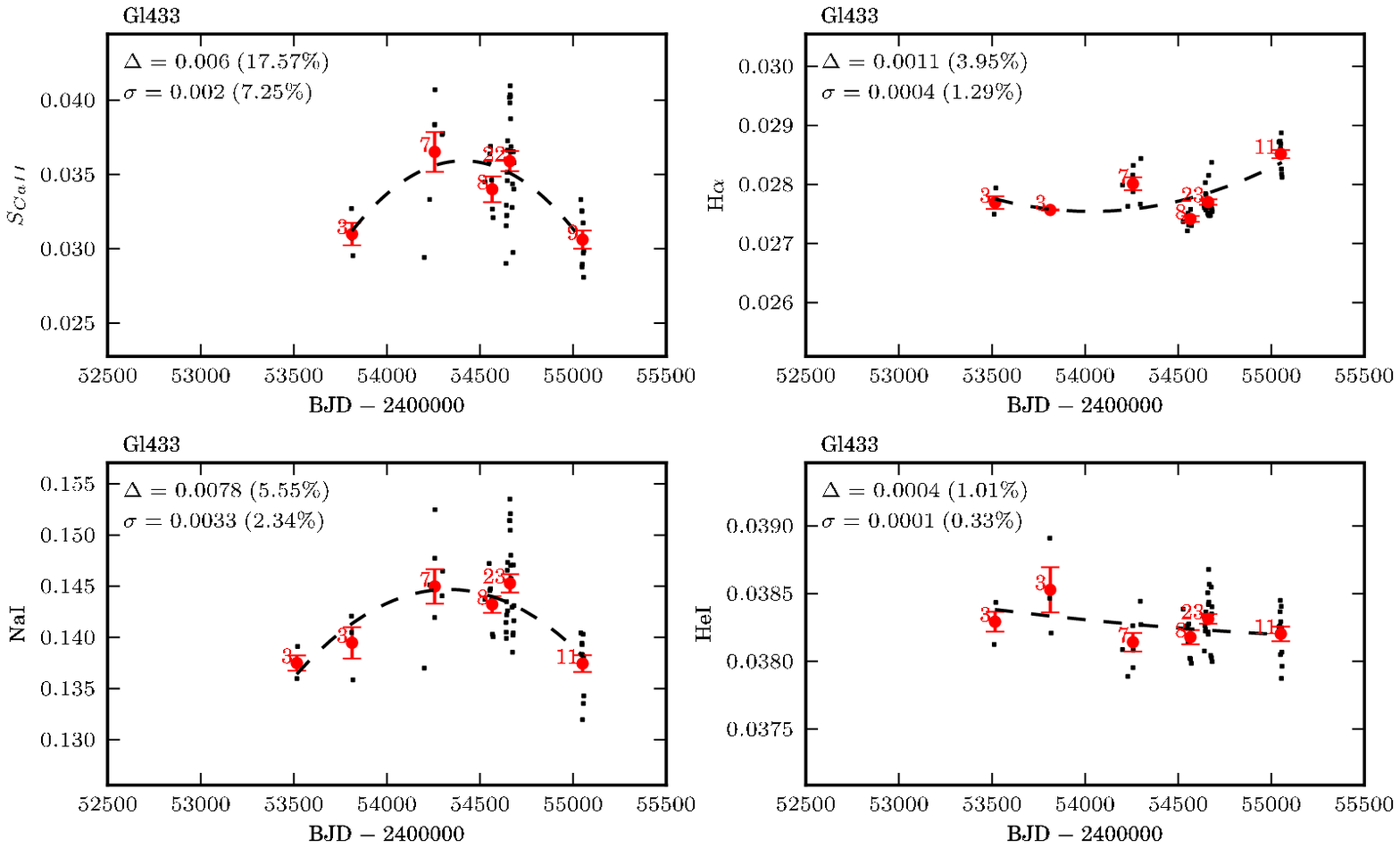}}
\resizebox{\hsize}{!}{\includegraphics{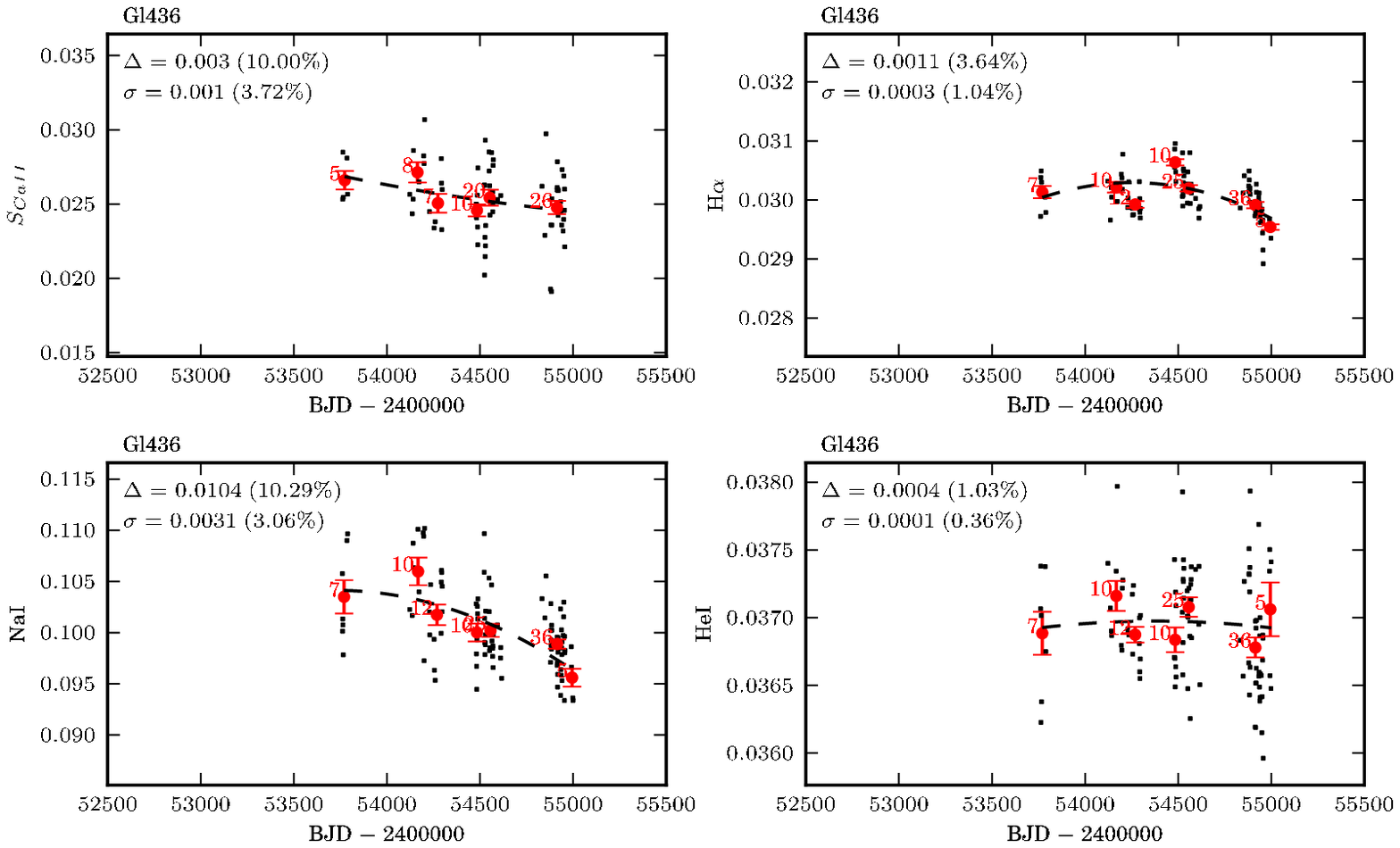}}
\caption{Continued from Fig. \ref{ind_bjd1}.}
\label{ind_bjd2}
\end{center}
\end{figure*}

\begin{figure*}[htbp]
\begin{center}
\resizebox{\hsize}{!}{\includegraphics{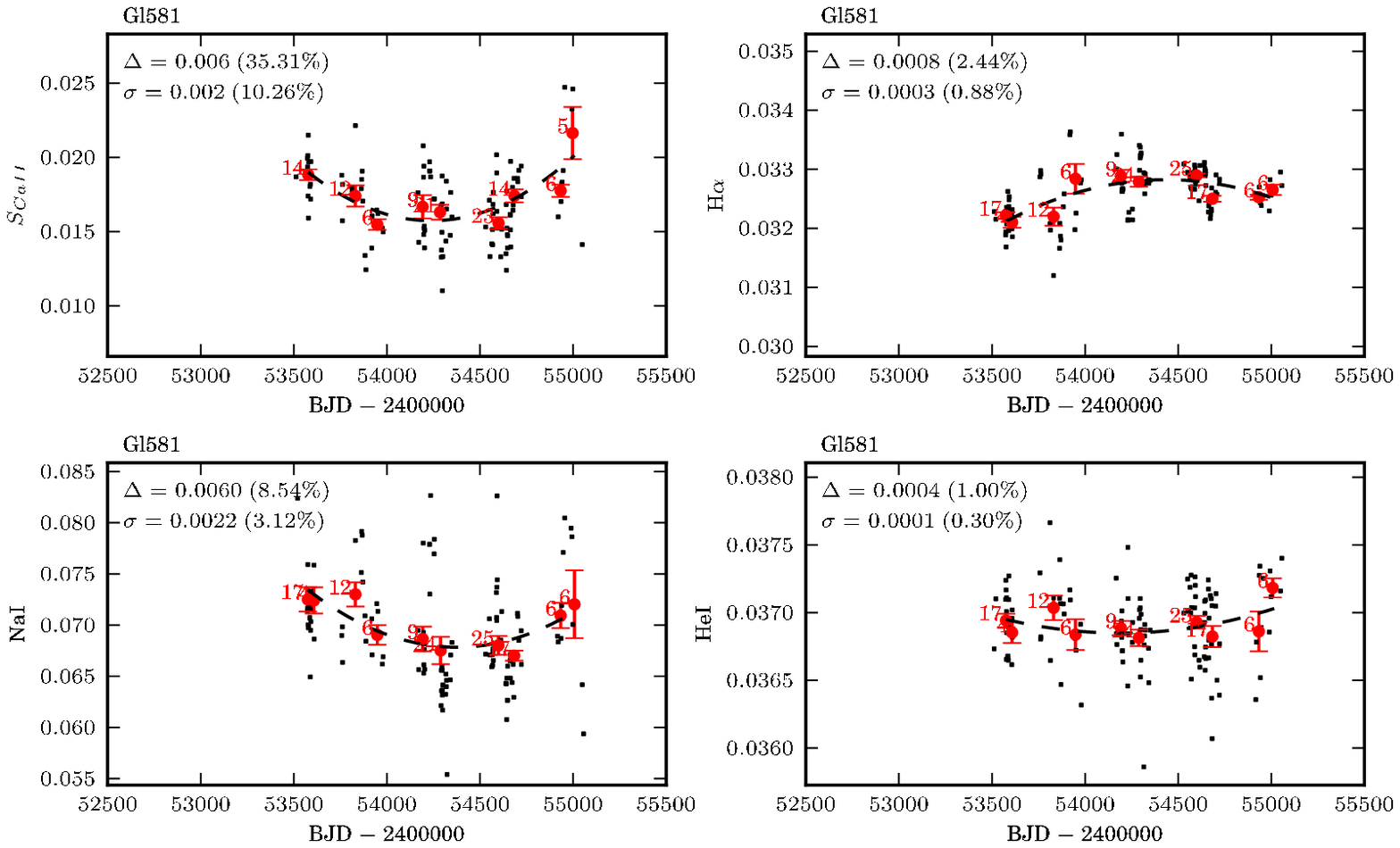}}
\resizebox{\hsize}{!}{\includegraphics{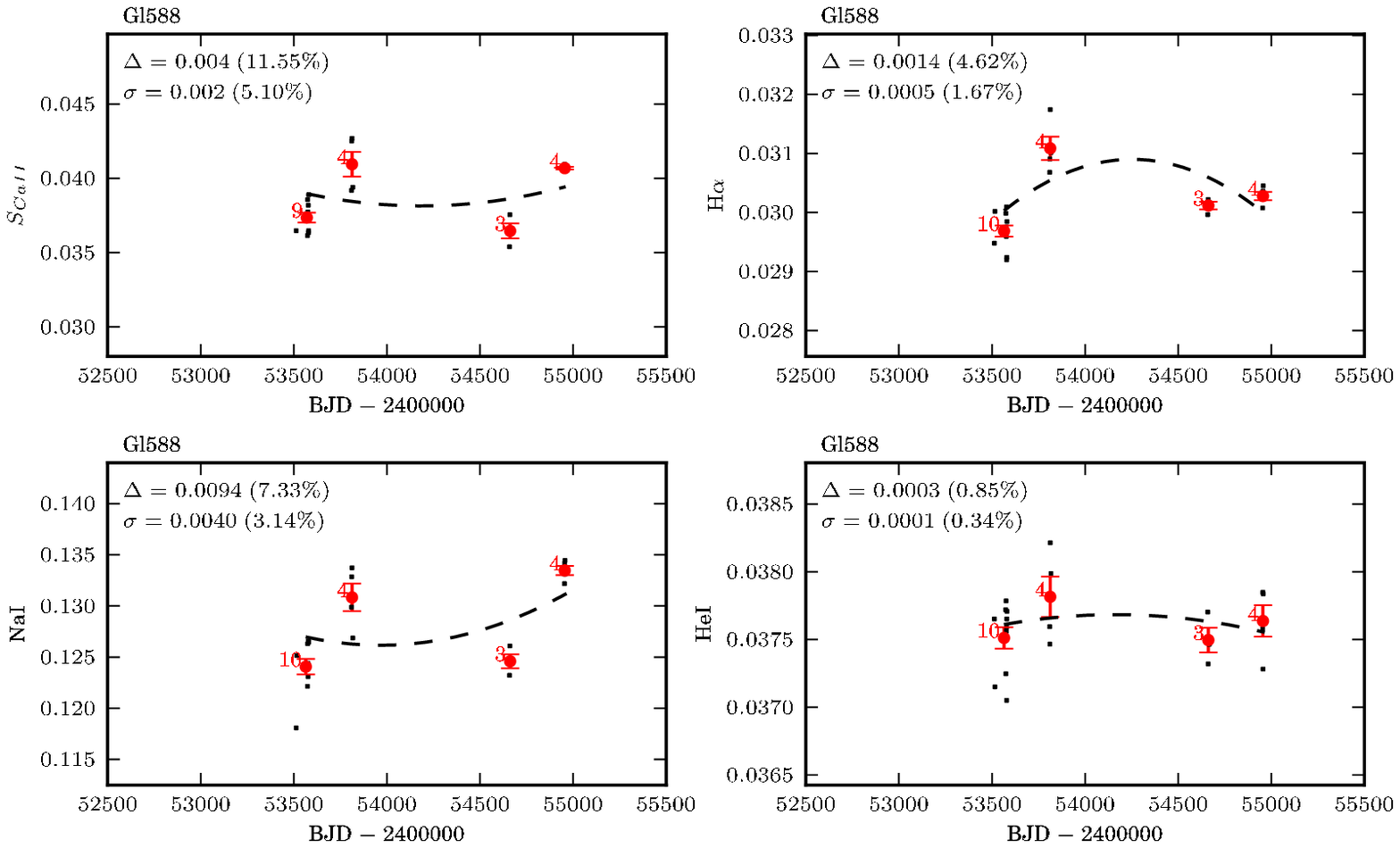}}
\caption{Continued from Fig. \ref{ind_bjd2}.}
\label{ind_bjd3}
\end{center}
\end{figure*}

\begin{figure*}[htbp]
\begin{center}
\resizebox{\hsize}{!}{\includegraphics{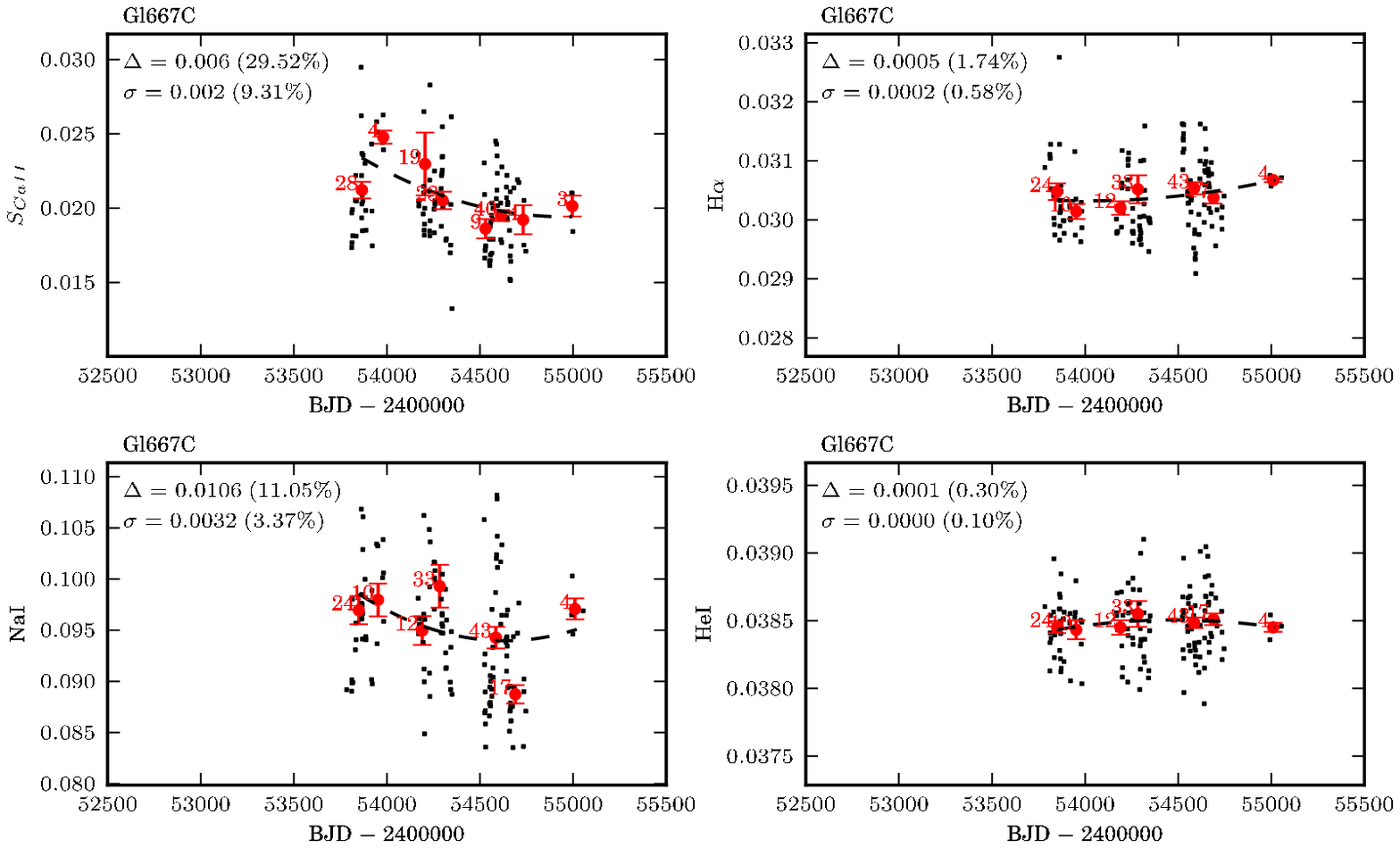}}
\resizebox{\hsize}{!}{\includegraphics{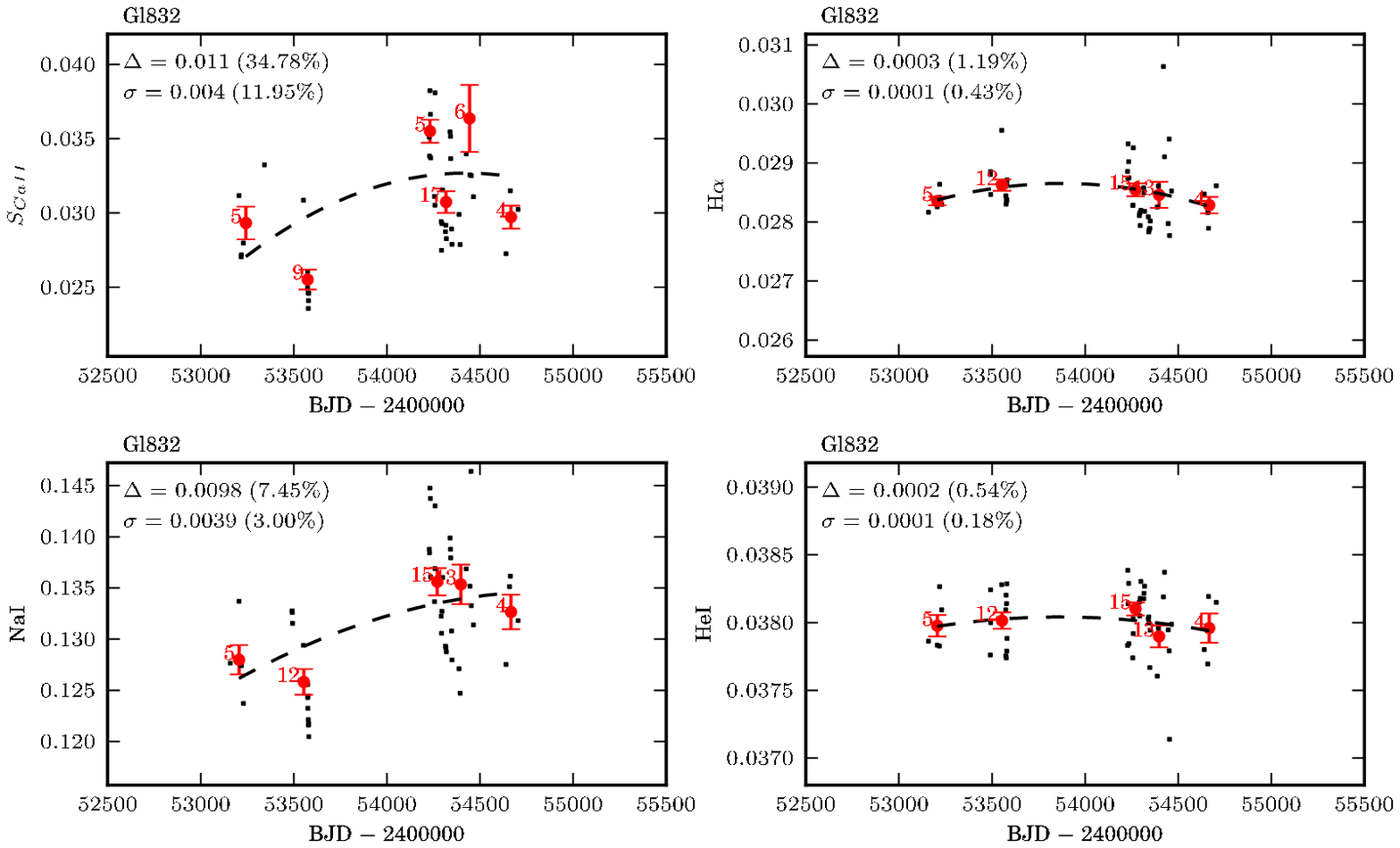}}
\caption{Continued from Fig. \ref{ind_bjd3}.}
\label{ind_bjd4}
\end{center}
\end{figure*}

\begin{figure*}[htbp]
\begin{center}
\resizebox{\hsize}{!}{\includegraphics{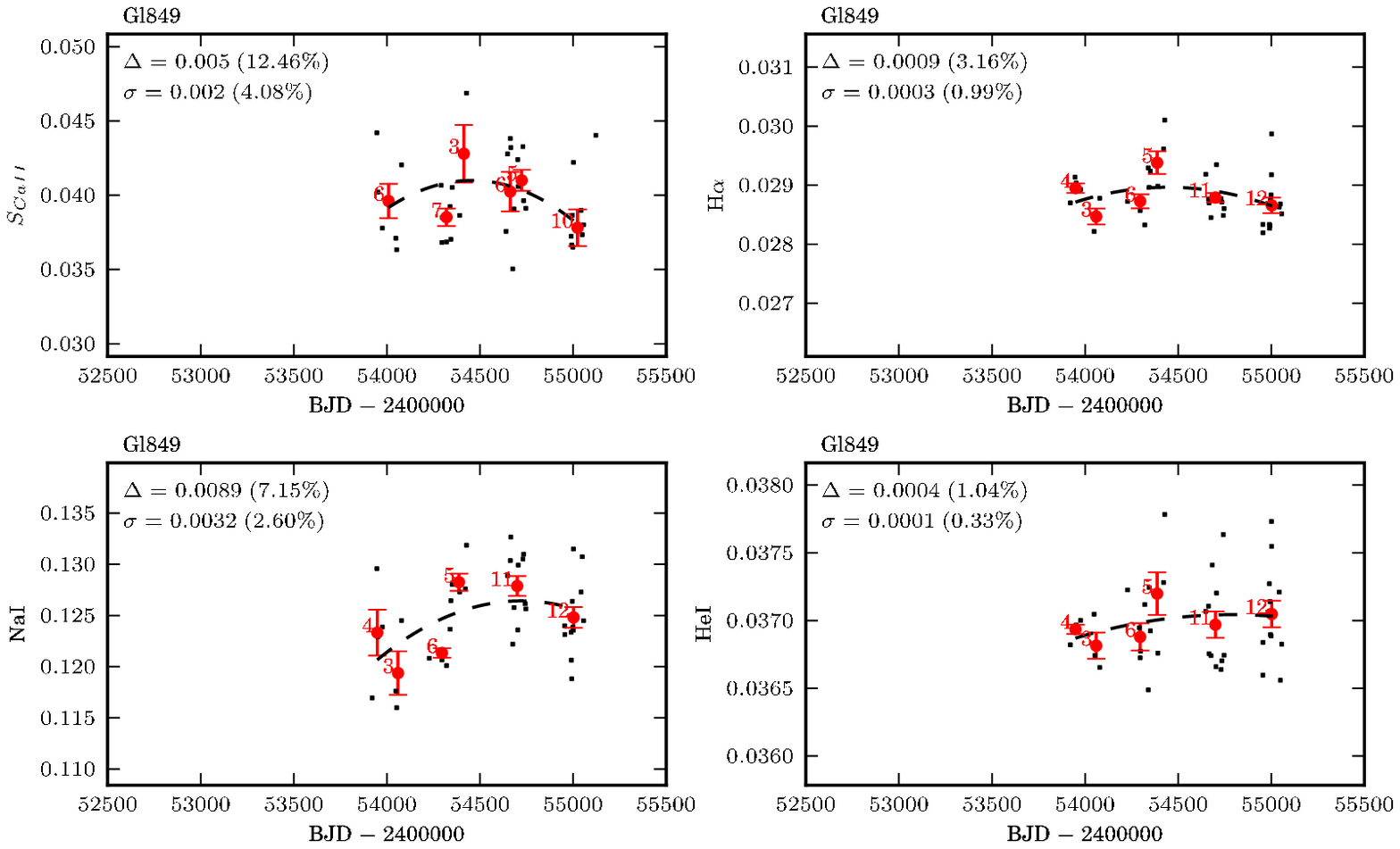}}
\resizebox{\hsize}{!}{\includegraphics{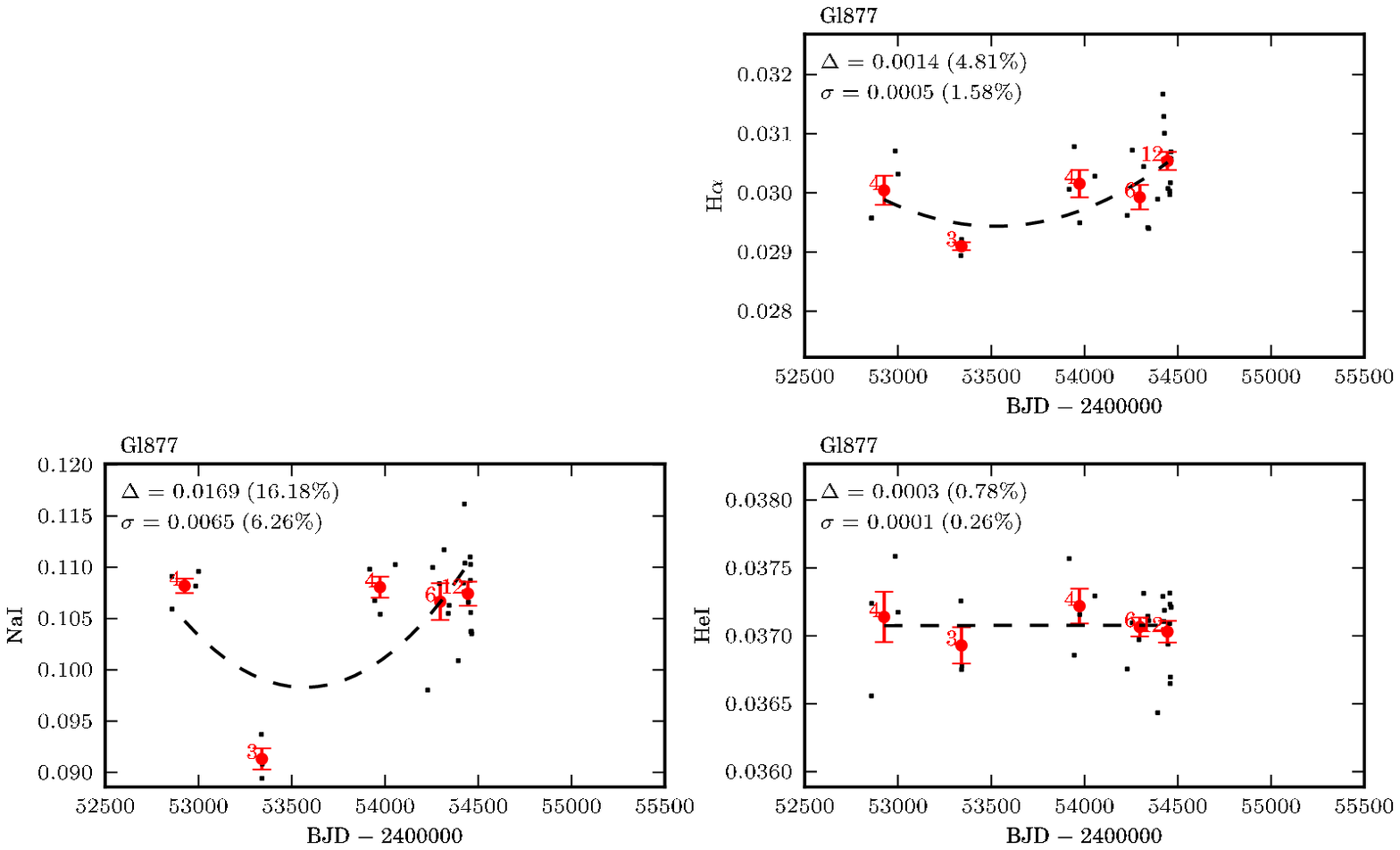}}
\caption{Continued from Fig. \ref{ind_bjd4}.}
\label{ind_bjd5}
\end{center}
\end{figure*}

\begin{figure*}[htbp]
\begin{center}
\resizebox{\hsize}{!}{\includegraphics{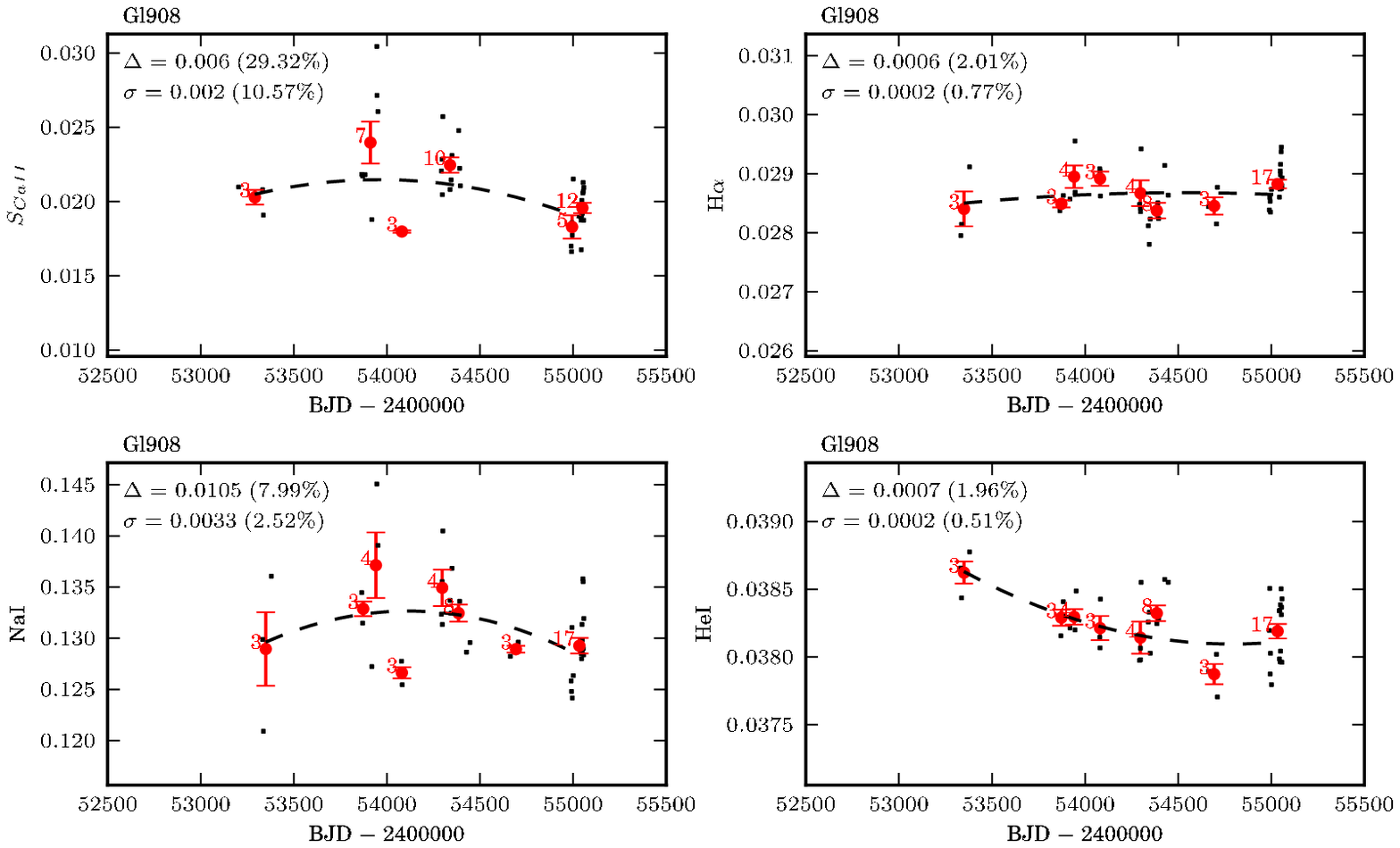}}
\resizebox{\hsize}{!}{\includegraphics{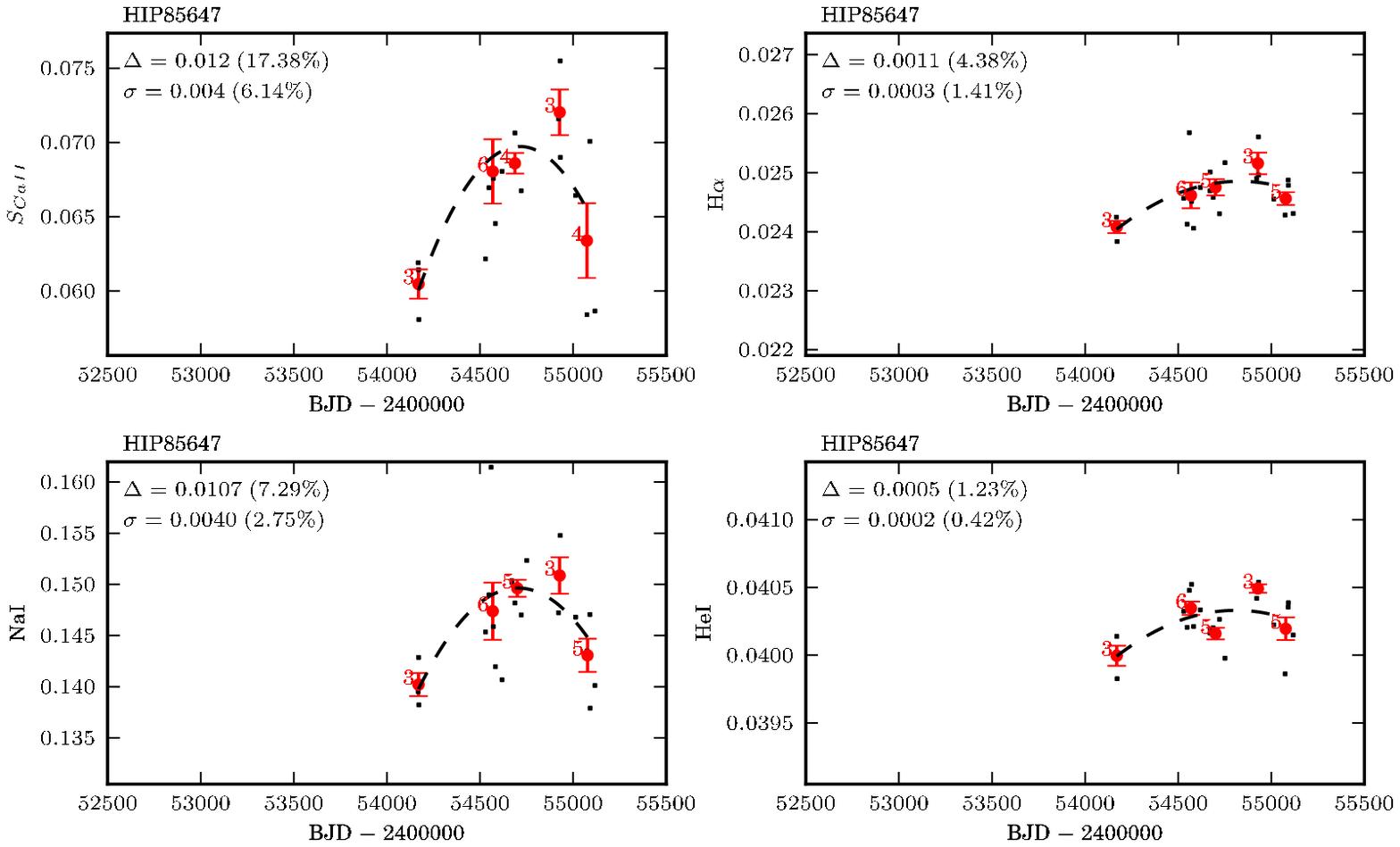}}
\caption{Continued from Fig. \ref{ind_bjd5}.}
\label{ind_bjd6}
\end{center}
\end{figure*}

\begin{landscape}
\begin{center}
\begin{table}[htbp] 
\caption{Variability F-tests for the activity indices. Bold values indicate probabilities lower than 0.05 (95\% significance level). The F-value used was $F = \sigma_e^2 / <\sigma_i>^2$ where $\sigma_e$ is the rms of the index (binned) and $<\sigma_i>$ the average of the error on the mean for the binned activity data of each star.}
\label{table:var}
\centering
\begin{tabular}{l c c |c c c| c c c| c c c| c c c} \\
\hline
\hline
\multicolumn{1}{l}{Star} &
\multicolumn{1}{l}{$N_{bins}(S)$} &
\multicolumn{1}{l}{$N_{bins}$} &
\multicolumn{3}{c}{$S_{\mathrm{Ca\,II}}$} &
\multicolumn{3}{c}{H$\alpha$} &
\multicolumn{3}{c}{\ion{Na}{i}} &
\multicolumn{3}{c}{\ion{He}{i}} \\
\multicolumn{1}{l}{} &
\multicolumn{1}{l}{} &
\multicolumn{1}{c}{} &
\multicolumn{1}{c}{$\sigma_e$} &
\multicolumn{1}{c}{$<\sigma_i>$} &
\multicolumn{1}{c}{$P(F_{const})$} &
\multicolumn{1}{c}{$\sigma_e$} &
\multicolumn{1}{c}{$<\sigma_i>$} &
\multicolumn{1}{c}{$P(F_{const})$} &
\multicolumn{1}{c}{$\sigma_e$} &
\multicolumn{1}{c}{$<\sigma_i>$} &
\multicolumn{1}{c}{$P(F_{const})$} &
\multicolumn{1}{c}{$\sigma_e$} &
\multicolumn{1}{c}{$<\sigma_i>$} &
\multicolumn{1}{c}{$P(F_{const})$} \\
\hline
GJ361	&	4	&	4	&	0.0028	&	0.0012	&	0.087		&	0.00034	&	0.00020	&	0.21			&	0.0032	&	0.0014	&	0.10			&	0.000079	&	0.00014	&	0.82 \\
GJ2049	&	$<4$	&	4	&			&			&				&	0.00019	&	0.00011	&	0.22			&	0.0036	&	0.0013	&	0.059		&	0.00014	&	0.000068	&	0.13 \\
GJ3218	&	$<4$	&	4	&			&			&				&	0.0013	&	0.00045	&	0.055		&	0.0077	&	0.0030	&	0.077		&	0.00019	&	0.00016	&	0.40 \\
Gl1		&	5	&	5	&	0.0015	&	0.00054	&	\textbf{0.034}	&	0.00038	&	0.00011	&	\textbf{0.018}	&	0.0018	&	0.0011	&	0.18			&	0.000081	&	0.00010	&	0.66 \\
Gl176	&	6	&	7	&	0.0036	&	0.0027	&	0.27			&	0.00061	&	0.00043	&	0.21			&	0.0064	&	0.0038	&	0.12			&	0.00027	&	0.00015	&	0.097 \\
Gl205	&	6	&	6	&	0.0031	&	0.0026	&	0.35			&	0.00029	&	0.00029	&	0.49			&	0.0045	&	0.0029	&	0.17			&	0.00018	&	0.000060	&	\textbf{0.015} \\
Gl273	&	5	&	6	&	0.0037	&	0.0010	&	\textbf{0.016}	&	0.00043	&	0.00022	&	0.080		&	0.0033	&	0.0014	&	\textbf{0.044}	&	0.000055	&	0.00014	&	0.97 \\
Gl382	&	5	&	5	&	0.0035	&	0.0031	&	0.41			&	0.00075	&	0.00043	&	0.15			&	0.0051	&	0.0027	&	0.12			&	0.000049	&	0.000091	&	0.87 \\
Gl393	&	$<4$	&	4	&			&			&				&	0.00028	&	0.00023	&	0.36			&	0.0020	&	0.0016	&	0.35			&	0.00016	&	0.00015	&	0.49 \\
Gl433	&	5	&	6	&	0.0024	&	0.00085	&	\textbf{0.033}	&	0.00036	&	0.000065	&	\textbf{0.00093}&	0.0033	&	0.0011	&	\textbf{0.014}	&	0.00013	&	0.000075	&	0.13 \\
Gl436	&	6	&	7	&	0.00095	&	0.00056	&	0.13			&	0.00031	&	0.000065	&	\textbf{0.00068}&	0.0031	&	0.00097	&	\textbf{0.0064}	&	0.00013	&	0.00011	&	0.32 \\
Gl479	&	4	&	4	&	0.0035	&	0.0024	&	0.28			&	0.00045	&	0.00047	&	0.53			&	0.0027	&	0.0026	&	0.48			&	0.00023	&	0.00014	&	0.22 \\
Gl526	&	4	&	4	&	0.0015	&	0.00077	&	0.15			&	0.00015	&	0.00011	&	0.31			&	0.0036	&	0.0011	&	\textbf{0.041}	&	0.000054	&	0.000075	&	0.71 \\
Gl551	&	$<4$	&	7	&			&			&				&	0.014	&	0.0078	&	0.11			&	0.043	&	0.046	&	0.56			&	0.0064	&	0.0035	&	0.081 \\
Gl581	&	9	&	10	&	0.0018	&	0.00064	&	\textbf{0.0041}	&	0.00029	&	0.000096	&	\textbf{0.0016}	&	0.0022	&	0.0013	&	0.072		&	0.00011	&	0.000079	&	0.17 \\
Gl588	&	4	&	4	&	0.0020	&	0.00044	&	\textbf{0.017}	&	0.00051	&	0.00011	&	\textbf{0.015}	&	0.0040	&	0.00081	&	\textbf{0.013}	&	0.00013	&	0.00011	&	0.40 \\
Gl667C	&	8	&	7	&	0.0019	&	0.00080	&	\textbf{0.016}	&	0.00018	&	0.00012	&	0.20			&	0.0032	&	0.0014	&	\textbf{0.026}	&	0.000039	&	0.000054	&	0.78 \\
Gl674	&	4	&	5	&	0.0012	&	0.0030	&	0.92			&	0.0012	&	0.00070	&	0.16			&	0.0037	&	0.0029	&	0.33			&	0.00025	&	0.00015	&	0.18 \\
Gl680	&	4	&	4	&	0.0019	&	0.0015	&	0.37			&	0.00019	&	0.00014	&	0.31			&	0.0024	&	0.0018	&	0.31			&	0.00028	&	0.00012	&	0.096 \\
Gl699	&	4	&	4	&	0.0011	&	0.0015	&	0.68			&	0.00065	&	0.00019	&	\textbf{0.038}	&	0.0022	&	0.0025	&	0.58			&	0.00019	&	0.00015	&	0.37 \\
Gl832	&	6	&	5	&	0.0037	&	0.0011	&	\textbf{0.0074}	&	0.00012	&	0.00013	&	0.53			&	0.0039	&	0.0015	&	\textbf{0.047}	&	0.000067	&	0.000075	&	0.58 \\
Gl849	&	6	&	6	&	0.0016	&	0.0012	&	0.23			&	0.00029	&	0.00012	&	\textbf{0.044}	&	0.0032	&	0.0013	&	\textbf{0.032}	&	0.00012	&	0.000097	&	0.31 \\
Gl876	&	4	&	4	&	0.0036	&	0.0014	&	0.075		&	0.00012	&	0.00014	&	0.63			&	0.0054	&	0.0017	&	\textbf{0.043}	&	0.00015	&	0.000082	&	0.18 \\
Gl877	&	$<4$	&	5	&			&			&				&	0.00047	&	0.00018	&	\textbf{0.044}	&	0.0065	&	0.0011	&	\textbf{0.0026}	&	0.000098	&	0.00012	&	0.65 \\
Gl887	&	4	&	4	&	0.0029	&	0.0011	&	0.077		&	0.00033	&	0.00015	&	0.11			&	0.0039	&	0.0014	&	0.063		&	0.00016	&	0.000067	&	0.088 \\
Gl908	&	6	&	8	&	0.0022	&	0.00060	&	\textbf{0.0071}	&	0.00022	&	0.00015	&	0.18			&	0.0033	&	0.0015	&	\textbf{0.024}	&	0.00020	&	0.000074	&	\textbf{0.010} \\
HIP12961	&	6	&	6	&	0.0015	&	0.0021	&	0.75			&	0.00017	&	0.00019	&	0.62			&	0.0029	&	0.0016	&	0.10			&	0.00013	&	0.000062	&	0.070 \\
HIP19394	&	$<4$	&	6	&			&			&				&	0.00031	&	0.00012	&	\textbf{0.031}	&	0.0036	&	0.0021	&	0.12			&	0.00019	&	0.00020	&	0.53 \\
HIP38594	&	$<4$	&	4	&			&			&				&	0.00040	&	0.000063	&	\textbf{0.0064}	&	0.0058	&	0.0020	&	0.055		&	0.00013	&	0.00012	&	0.42 \\
HIP85647	&	5	&	5	&	0.0041	&	0.0016	&	\textbf{0.046}	&	0.00035	&	0.00015	&	0.067		&	0.0040	&	0.0016	&	0.055		&	0.00017	&	0.000056	&	\textbf{0.028} \\
\hline
\end{tabular}
\end{table}
\end{center}
\end{landscape}

\subsection{$S_{\mathrm{Ca\,II}}$ variability}
Although the spectral orders of the \ion{Ca}{ii} H \& K lines are the ones with the lower signal-to-noise ratio, the $S_{\mathrm{Ca\,II}}$ index still shows a great number of stars with variability. From our sample of 23 stars, 9 have P-values lower than or equal to 0.05, which represents 39\% of the sample. These stars are Gl1, Gl273, Gl433, Gl581, Gl588, Gl667C, Gl832, Gl908, and HIP85647. If we include the three stars with P-value below 0.1 (GJ361, Gl876, and Gl887; in general they have short observation timespans or only a few nights and thus any variations caused by their hypothetic cycle might be underestimated) we have 52\% of the sample with significant mid-term activity variations. Moreover, all stars with significant variations (P $\leq$ 0.05) in the $S_{\mathrm{Ca\,II}}$ index also present significant variations in at least one of the other indices, which reinforces this variable behavior.

\subsection{$\mathrm{H\alpha}$ variability}
The H$\alpha$ index indicates significant variability (P $\leq$ 0.05) for 10 stars (Gl1, Gl433, Gl436, Gl581, Gl588, Gl699, Gl849, Gl877, HIP19394, and HIP38594) out of 30, representing 33\% of the sample. All these stars, apart from Gl699, HIP19394 and HIP38594, also have significant variations in the other indices.
If we include the three stars with a P-values below 0.1 (GJ3218, Gl273, and HIP85647), then around 43\% of the sample shows variability in this index, and two of these stars present a strong variability in the $S_{\mathrm{Ca\,II}}$ and  \ion{Na}{i} indices.

\subsection{$\ion{Na}{i}$ variability}
In terms of variability, the \ion{Na}{i} index seems to be one of the best we have. Out of 30 stars, 11 present significant scatter (Gl273, Gl433, Gl436, Gl526, Gl588, Gl667C, Gl832, Gl849, Gl876, Gl877, and Gl908, representing 37\% of the sample), and there are six more with 0.05 $<$ P $\leq$ 0.1 (GJ361, GJ2049, GJ3218, Gl581, HIP38594, and HIP85647), which, if these are included, amounts to 57\% of the sample. Apart from Gl526, all stars with P $\leq$ 0.05 also have a strong variability or hints of variability in the other indices.

\subsection{$\ion{He}{i}$ variability}
The \ion{He}{i} index is the one with fewer stars presenting variability in the F-tests. Only three stars in the sample (Gl205, Gl908 and HIP85647, 10\% of the sample) have P $\leq$ 0.05 and five (Gl176, Gl551, Gl680, Gl887, and HIP12961) have 0.05 $<$ P $\leq$ 0.1, which yields an overall of 27\% of the sample with variability. This indicates that this index, although in a region of the spectrum with high signal-to-noise ratio, does not vary with a sufficiently high amplitude to be detected in M dwarfs. It is curious that Gl205, although having a strong variability in the \ion{He}{i} index, does not present any significant scatter either in $S_{\mathrm{Ca\,II}}$ (P = 0.35) or H$\alpha$ (P = 0.49). The same is true of Gl176 and HIP12961. On the other hand, there are a large number of stars that have strong variations in the other indices, but their P-values for \ion{He}{i} indicate no variability. Therefore, this index does not seem to be a good proxy of medium to long-term activity variations of early-M type dwarfs.

\section{Activity maxima and minima}\label{trends}
Apart from the variability tests we also performed F-tests to try to find maxima or minima in our data that could indicate the presence of magnetic cycles. We fitted straight lines and second-order polynomials to the time-series and calculated an F-test using $F_{poly} = (N - 3)(\chi^2_{slope} - \chi^2_{poly})/\chi^2_{poly}$, where $\chi^2_{slope}$ is the chi-squared of the straight line model, $\chi^2_{poly}$ the chi-squared of the polynomial model, and $N$ the number of bins. The probability that the data are better fitted by a polynomial than by a straight line is given in Table \ref{ftest} by low values of $P(F_{poly})$. We used P-values lower than 0.05 as a limit for statistically significant improvements to the fit.

The $S_{\mathrm{Ca\,II}}$ index shows three stars having maxima or minima, namely Gl1, Gl433, and Gl581, whose data are statistically better fitted by a second-order polynomial than by a straight line. Both Gl1 and Gl433 show a maximum of activity in the time-series plots (Figures \ref{ind_bjd1} and \ref{ind_bjd2}), while Gl581 presents a minimum (Figure \ref{ind_bjd3}).

The $\ion{Na}{i}$ index shows the same behavior in the tests as the calcium lines apart from HIP85647. This star passed the test and shows a maximum in the time-series (Figure \ref{ind_bjd6}) with a similar behavior also observed on $S_{\mathrm{Ca\,II}}$ (P = 0.11) and H$\alpha$ (P = 0.12).
Two other stars passed the test for the H$\alpha$ index, Gl832 and HIP38594, but without similar variations observed for the other activity indicators. For Gl581 we observe a maximum instead of the minimum produced by the $\ion{Ca}{ii}$ and $\ion{Na}{ii}$ lines.

Two stars present significant maximum or minimum in \ion{He}{i}. These are Gl361, which has passed the test for the other three indices, and Gl581 with a minimum.

\begin{center}
\begin{table*}[tbp]
\caption{F-tests for trends in our activity indices. The F-value used was $F_{poly} = (N - 3)(\chi^2_{slope} - \chi^2_{poly})/\chi^2_{poly}$, where $\chi^2_{slope}$ is the chi-squared of the straight line model, $\chi^2_{poly}$ the chi-squared of the polynomial model, and $N$ the number of bins. This test will give the probability that a second-order polynomial is a better fit to the data than a straight line. Bold values indicate probabilities lower than 0.05 (95\% significance level).}
\label{ftest}
\centering
\begin{tabular}{l c c | c c | c c | c c | c c} \\
\hline
\hline
\multicolumn{1}{l}{Star} &
\multicolumn{1}{l}{$N_{bins}(S)$} &
\multicolumn{1}{l}{$N_{bins}$} &
\multicolumn{2}{c}{$S_{\mathrm{Ca\,II}}$} &
\multicolumn{2}{c}{H$\alpha$} &
\multicolumn{2}{c}{\ion{Na}{i}} &
\multicolumn{2}{c}{\ion{He}{i}} \\
\multicolumn{1}{l}{} &
\multicolumn{1}{c}{} &
\multicolumn{1}{c}{} &
\multicolumn{1}{c}{$F_{poly}$} &
\multicolumn{1}{c}{$P(F_{poly})$} &
\multicolumn{1}{c}{$F_{poly}$} &
\multicolumn{1}{c}{$P(F_{poly})$} &
\multicolumn{1}{c}{$F_{poly}$} &
\multicolumn{1}{c}{$P(F_{poly})$} &
\multicolumn{1}{c}{$F_{poly}$} &
\multicolumn{1}{c}{$P(F_{poly})$} \\
\hline
GJ361	&	4	&	4	&	0.475	&	0.72			&	3.13		&	0.37			&	0.00785	&	0.99			&	414		&	\textbf{0.035} \\
GJ2049	&	$<4$	&	4	&			&				&	0.187	&	0.85			&	0.0204	&	0.98			&	0.0733	&	0.93 \\
GJ3218	&	$<4$	&	4	&			&				&	5.44		&	0.29			&	5.46		&	0.29			&	0.00831	&	0.99 \\
Gl1		&	5	&	5	&	27.9		&	\textbf{0.019}	&	3.61		&	0.12			&	22.5		&	\textbf{0.023}	&	0.729	&	0.34 \\
Gl176	&	6	&	7	&	1.33		&	0.24			&	0.579	&	0.41			&	0.442	&	0.46			&	0.827	&	0.34 \\
Gl205	&	6	&	6	&	0.00533	&	0.56			&	0.0243	&	0.56			&	0.166	&	0.53			&	0.0851	&	0.55 \\
Gl273	&	5	&	6	&	0.0998	&	0.52			&	0.0967	&	0.55			&	5.18		&	0.059		&	0.0761	&	0.55 \\
Gl382	&	5	&	5	&	0.511	&	0.39			&	0.820	&	0.32			&	0.150	&	0.50			&	0.136	&	0.51 \\
Gl393	&	$<4$	&	4	&			&				&	0.0149	&	0.99			&	27.6		&	0.13			&	0.0533	&	0.95 \\
Gl433	&	5	&	6	&	17.6		&	\textbf{0.030}	&	5.15		&	0.059		&	18.4		&	\textbf{0.011}	&	0.0255	&	0.56 \\
Gl436	&	6	&	7	&	0.196	&	0.52			&	2.79		&	0.098		&	0.769	&	0.35			&	0.111	&	0.56 \\
Gl479	&	4	&	4	&	24.1		&	0.14			&	6.86		&	0.26			&	2.92		&	0.38			&	10.9		&	0.21 \\
Gl526	&	4	&	4	&	6.02		&	0.28			&	0.666	&	0.66			&	9.94		&	0.22			&	0.168	&	0.87 \\
Gl551	&	$<4$	&	7	&			&				&	0.193	&	0.54			&	0.795	&	0.35			&	0.510	&	0.44 \\
Gl581	&	9	&	10	&	11.9		&	\textbf{0.0022}	&	12.2		&	\textbf{0.0010}	&	10.7		&	\textbf{0.0016}	&	4.45		&	\textbf{0.019} \\
Gl588	&	4	&	4	&	0.469	&	0.72			&	0.290	&	0.80			&	0.322	&	0.78			&	0.0387	&	0.96 \\
Gl667C	&	8	&	7	&	0.713	&	0.38			&	3.55		&	0.070		&	1.26		&	0.24			&	4.06		&	0.057 \\
Gl674	&	4	&	5	&	0.128	&	0.89			&	0.826	&	0.32			&	0.467	&	0.40			&	0.0445	&	0.54 \\
Gl680	&	4	&	4	&	0.0706	&	0.94			&	1.48		&	0.50			&	4.78		&	0.31			&	0.905	&	0.60 \\ 
Gl699	&	4	&	4	&	2.92		&	0.38			&	62.9		&	0.089		&	1.74		&	0.47			&	0.0186	&	0.98 \\
Gl832	&	6	&	5	&	0.153	&	0.54			&	63.7		&	\textbf{0.0084}	&	0.0227	&	0.54			&	0.996	&	0.29 \\
Gl849	&	6	&	6	&	0.285	&	0.49			&	0.673	&	0.37			&	0.325	&	0.48			&	1.08		&	0.28 \\
Gl876	&	4	&	4	&	1.08		&	0.56			&	4.78		&	0.31			&	1.63		&	0.48			&	6.21		&	0.27 \\
Gl877	&	$<4$	&	5	&			&				&	3.56		&	0.12			&	1.41		&	0.24			&	0.00635	&	0.54 \\
Gl887	&	4	&	4	&	9.44		&	0.22			&	0.465	&	0.72			&	0.756	&	0.63			&	0.00450	&	0.99 \\
Gl908	&	6	&	8	&	1.16		&	0.27			&	0.689	&	0.39			&	0.970	&	0.30			&	2.29		&	0.11 \\
HIP12961	&	6	&	6	&	0.0851	&	0.55			&	0.136	&	0.54			&	2.15		&	0.16			&	0.451	&	0.43 \\
HIP19394	&	$<4$	&	6	&			&				&	1.37		&	0.23			&	2.15		&	0.16			&	5.83		&	0.051 \\
HIP38594	&	$<4$	&	4	&			&				&	1399		&	\textbf{0.019}	&	37.3		&	0.12			&	0.322	&	0.78 \\
HIP85647	&	5	&	5	&	4.15		&	0.11			&	3.66		&	0.12			&	15.3		&	\textbf{0.034}	&	0.133	&	0.51 \\
\hline
\end{tabular}
\end{table*}
\end{center}

\section{Comparision between activity indicators}\label{corr}
After assessing the variability for our four indices, we now compare them to assess their mid-term correlations.
Table \ref{table:ind} shows the Pearson correlation coefficients between the activity indices for our sample ($\rho$). $N_{bins}$ is the number of bins for each star ($N_{bins}(S)$ for comparisons using the $S$ index). The false-alarm probability (FAP) was computed by bootstraping the nightly averaged data, then binning the data every 150 days, and calculating the coefficient for each of the 10\,000 permutations. A significant FAP was then chosen for values $< 0.05$ (95\% significance level) and highlighted in bold.

\begin{center}
\begin{table*}[htbp]
\caption{Pearson correlation coefficients between the activity indices. FAPs calculated using bootstrap permutations (see text). In bold are given FAP values below 0.05. $N_{bins}$ is the number of bins for each star.}
\label{table:ind}
\centering
\begin{tabular}{l cc | cc | cc | cc | cc | cc | cc} \\
\hline
\hline
\multicolumn{1}{l}{Star} &
\multicolumn{1}{l}{$N_{bins}(S)$} &
\multicolumn{1}{l}{$N_{bins}$} &
\multicolumn{2}{c}{$S_{\mathrm{Ca\,II}}$ vs H$\alpha$} &
\multicolumn{2}{c}{$S_{\mathrm{Ca\,II}}$ vs \ion{Na}{i}} &
\multicolumn{2}{c}{$S_{\mathrm{Ca\,II}}$ vs \ion{He}{i}} &
\multicolumn{2}{c}{H$\alpha$ vs \ion{Na}{i}} &
\multicolumn{2}{c}{H$\alpha$ vs \ion{He}{i}} &
\multicolumn{2}{c}{\ion{Na}{i} vs \ion{He}{i}} \\
\multicolumn{1}{l}{} &
\multicolumn{1}{c}{} &
\multicolumn{1}{c}{} &
\multicolumn{1}{c}{$\rho$} &
\multicolumn{1}{c}{FAP} &
\multicolumn{1}{c}{$\rho$} &
\multicolumn{1}{c}{FAP} &
\multicolumn{1}{c}{$\rho$} &
\multicolumn{1}{c}{FAP} &
\multicolumn{1}{c}{$\rho$} &
\multicolumn{1}{c}{FAP} &
\multicolumn{1}{c}{$\rho$} &
\multicolumn{1}{c}{FAP} &
\multicolumn{1}{c}{$\rho$} &
\multicolumn{1}{c}{FAP} \\
\hline
GJ361	&	4	&	4	&	0.58	&	0.27			&	0.73	&	0.12			&	0.43	&	0.35			&	-0.07	&	0.46			&	0.86	&	0.13			&	-0.24	&	0.41 \\
GJ2049	&	$<4$	&	4	&		&				&		&				&		&				&	-0.04	&	0.47			&	-0.89	&	0.083		&	0.43	&	0.33 \\
GJ3218	&	$<4$	&	4	&		&				&		&				&		&				&	1.00	&	\textbf{0.0006}	&	0.39	&	0.36			&	0.36	&	0.37 \\
Gl1		&	5	&	5	&	0.12	&	0.41			&	0.89	&	\textbf{0.031}	&	0.56	&	0.24			&	0.37	&	0.33			&	0.14	&	0.45			&	0.84	&	0.067 \\
Gl176	&	6	&	7	&	0.85	&	\textbf{0.016}	&	0.33	&	0.27			&	0.76	&	\textbf{0.033}	&	0.40	&	0.24			&	0.77	&	\textbf{0.034}	&	-0.04	&	0.48 \\
Gl205	&	6	&	6	&	0.98	&	\textbf{0.0015}	&	0.94	&	\textbf{0.0079}	&	0.58	&	0.17			&	0.83	&	0.054		&	0.57	&	0.17			&	0.69	&	0.10 \\
Gl273	&	5	&	6	&	0.06	&	0.46			&	0.90	&	\textbf{0.024}	&	0.45	&	0.25			&	-0.40	&	0.24			&	0.80	&	\textbf{0.038}	&	0.08	&	0.44 \\
Gl382	&	5	&	5	&	0.99	&	\textbf{0.0007}	&	0.98	&	\textbf{0.0017}	&	0.42	&	0.24			&	0.98	&	\textbf{0.0021}	&	0.46	&	0.23			&	0.56	&	0.15 \\
Gl393	&	$<4$	&	4	&		&				&		&				&		&				&	-0.10	&	0.45			&	0.91	&	\textbf{0.044}	&	0.17	&	0.42 \\
Gl433	&	5	&	6	&	-0.28	&	0.32			&	0.98	&	\textbf{0.0012}	&	-0.50	&	0.29			&	-0.32	&	0.25			&	-0.39	&	0.30			&	-0.31	&	0.33 \\
Gl436	&	6	&	7	&	-0.06	&	0.45			&	0.87	&	\textbf{0.010}	&	0.42	&	0.20			&	0.49	&	0.19			&	-0.15	&	0.38			&	0.21	&	0.32 \\
Gl479	&	4	&	4	&	0.99	&	\textbf{0.0044}	&	0.94	&	\textbf{0.034}	&	0.87	&	0.060		&	0.94	&	\textbf{0.033}	&	0.82	&	0.10			&	0.91	&	\textbf{0.050} \\
Gl526	&	4	&	4	&	-0.78	&	0.13			&	0.99	&	\textbf{0.013}	&	0.21	&	0.36			&	-0.86	&	0.12			&	0.42	&	0.24			&	0.10	&	0.43 \\
Gl551	&	$<4$	&	7	&		&				&		&				&		&				&	0.59	&	0.086		&	0.96	&	\textbf{0.0007}	&	0.55	&	0.11 \\
Gl581	&	9	&	10	&	-0.44	&	0.094		&	0.82	&	\textbf{0.0036}	&	0.75	&	\textbf{0.024}	&	-0.75	&	\textbf{0.0035}	&	-0.14	&	0.37			&	0.58	&	0.063 \\
Gl588	&	4	&	4	&	0.73	&	0.11			&	0.94	&	\textbf{0.026}	&	0.88	&	0.067		&	0.63	&	0.17			&	0.93	&	\textbf{0.036}	&	0.73	&	0.15 \\ 
Gl667C	&	8	&	7	&	-0.31	&	0.23			&	0.89	&	\textbf{0.0016}	&	-0.04	&	0.40			&	0.11	&	0.40			&	0.35	&	0.23			&	-0.15	&	0.34 \\
Gl674	&	4	&	5	&	0.35	&	0.36			&	0.91	&	\textbf{0.045}	&	0.60	&	0.23			&	0.59	&	0.18			&	0.81	&	\textbf{0.036}	&	0.85	&	\textbf{0.028} \\
Gl680	&	4	&	4	&	0.96	&	\textbf{0.022}	&	0.85	&	0.085		&	0.89	&	0.087		&	0.95	&	\textbf{0.021}	&	0.13	&	0.39			&	-0.04	&	0.50 \\
Gl699	&	4	&	4	&	0.57	&	0.21			&	0.43	&	0.32			&	0.25	&	0.38			&	0.32	&	0.35			&	0.89	&	0.062		&	0.48	&	0.27 \\
Gl832	&	6	&	5	&	0.64	&	0.083		&	0.97	&	\textbf{0.0006}	&	0.05	&	0.47			&	-0.17	&	0.37			&	0.49	&	0.15			&	-0.02	&	0.49 \\
Gl849	&	6	&	6	&	0.80	&	0.063		&	0.71	&	0.069		&	0.44	&	0.24			&	0.71	&	0.077		&	0.82	&	\textbf{0.041}	&	0.84	&	\textbf{0.025} \\
Gl876	&	4	&	4	&	0.76	&	0.14			&	1.00	&	\textbf{0.0005}	&	0.58	&	0.24			&	0.88	&	0.054		&	0.88	&	0.053		&	0.93	&	\textbf{0.032} \\
Gl877	&	$<4$	&	5	&		&				&		&				&		&				&	0.90	&	\textbf{0.032}	&	0.57	&	0.20			&	0.79	&	0.088 \\
Gl887	&	4	&	4	&	0.85	&	0.12			&	0.91	&	0.063		&	0.63	&	0.21			&	0.98	&	\textbf{0.018}	&	0.82	&	0.10			&	0.86	&	0.082 \\
Gl908	&	6	&	8	&	-0.24	&	0.33			&	0.89	&	\textbf{0.016}	&	0.31	&	0.30			&	0.12	&	0.39			&	-0.16	&	0.37			&	0.07	&	0.44 \\
HIP12961	&	6	&	6	&	0.35	&	0.25			&	0.30	&	0.32			&	0.63	&	0.098		&	-0.38	&	0.23			&	0.83	&	\textbf{0.018}	&	-0.78	&	\textbf{0.033} \\
HIP19394	&	$<4$	&	6	&		&				&		&				&		&				&	0.39	&	0.21			&	0.08	&	0.45			&	0.26	&	0.31 \\
HIP38594	&	$<4$	&	4	&		&				&		&				&		&				&	0.59	&	0.19			&	0.19	&	0.41			&	-0.15	&	0.43 \\ 
HIP85647	&	5	&	5	&	0.94	&	\textbf{0.011}	&	0.99	&	\textbf{0.0016}	&	0.94	&	\textbf{0.012}	&	0.92	&	\textbf{0.019}	&	0.88	&	\textbf{0.037}	&	0.77	&	0.087 \\
\hline
\end{tabular}
\end{table*}
\end{center}

\subsection{$S_{\mathrm{Ca\,II}}$ vs $\mathrm{H\alpha}$} \label{s_ha}

Six stars (out of 23, corresponding to 26\% of the sample) show a significant positive correlation between the $S_{\mathrm{Ca\,II}}$ and H$\alpha$ indices (Table \ref{table:ind}). Around 57\% of the sample have a positive correlation between $S$ and H$\alpha$ higher than 0.5. Although it has been claimed that a strong correlation between these two indices exists, it is not clear if that is the case for all stellar types or levels of activity \citep[e.g.][]{montes1995,stressmeier1990,robinson1990,giampapa1989}.

\citet{cincunegui2007b} studied the correlation between these two indices for a sample of 109 stars ranging in spectral type from F6 to M5. These authors found a great variety of correlations between $-1$ and 1.

More recently, \cite{walkowicz2009} compared simultaneous spectra of \ion{Ca}{ii} and Balmer lines for a sample of M3 dwarfs and observed that the relationship between calcium and H$\alpha$ is not linear: weak absorption of the H$\alpha$ line can correspond either to weak or intermediate \ion{Ca}{ii} K emission, as proposed by \cite{cram1987} and previously observed by \cite{stauffer1986} \citep[see also][]{rauscher2006}.

\citet{meunier2009} also studied these indices for the Sun, and although we used 150 day bins and they used individual measurements, their discussion about the relation between $\ion{Ca}{ii}$ and H$\alpha$ can provide interesting insights here. These authors pointed out that if the timescale of the observations covered less than the solar cycle, the correlation could decrease to negative values. We do not have information about the length of the cycles of these stars (or if they have periodic cycles), but because our maximum timespans are in the 6-year range or less, the conclusion by \citet{meunier2009} for the low positive or negative values of the correlation coefficient might be a possibility. Another possibility to explain the low and negative values of the correlation coefficients might be explained by the different sensitivity of the indices to different activity phenomena. For the Sun, the surface coverage of plages as measured by the two indices is not exactly the same: it is smaller for H$\alpha$ than for the $\ion{Ca}{ii}$ lines \citep{meunier2009}. This will reduce the correlation between them. Furthermore, the H$\alpha$ line is more sensitive to the presence of dark filaments. As \citet{meunier2009} pointed out, this will reduce the correlation coefficient to values closer to zero if the filaments and plages are not sharing the same positions on the disk of the star. If the filaments are spatially well-correlated with plages, the correlation between the two indices will tend to negative values. Thus, the anti-correlation found for Gl526 with a Pearson correlation coefficient of $-0.78$ (FAP = 13\%) can be an indication that the plages and filaments are spatially well-correlated for that star. Stars with strong positive correlations (and higher levels of activity) such as Gl176, Gl205, Gl382, and Gl479 probably have a saturated filament contribution to H$\alpha$ that therefore does not influence the index as much as the plage contribution.

Figure \ref{r_s_ha} (left panel) shows the Pearson correlation coefficient between these two indices against the mean values of the $S$-index. Open squares are values with FAP $\leq 0.01$, open circles for $0.01 <$ FAP $\leq 0.05$, and "plus" ("$+$") symbols for FAP values higher than 0.05.

As was found by \citet{cincunegui2007b}, we obtain a great variety of correlations from around -0.8 to 1. Nevertheless, there is a tendency in our sample for the positive correlations. A trend can be observed when the correlation coefficient is plotted against the mean $S_{\mathrm{Ca\,II}}$: for values of $S$ lower than around  0.035 there are no significant correlations and for $S$ higher than $\sim$0.035, there are only positive correlations and some of them with statistical significance. This trend is not observed when the correlation is plotted against the mean values of H$\alpha$.

An interpretation of this could be that because the H$\alpha$ is more sensitive to filaments than the $S$-index, as the activity gets stronger (higher $S_{\mathrm{Ca\,II}}$) the contribution of plages becomes more important to the H$\alpha$ index than the contribution coming from filaments, because their contribution saturates at a certain activity level \citep{meunier2009}. This will produce the observed positive correlation between the two indices for higher values of $S$. 
We could also argue that this is an effect of the difficulty of obtaining statistically significant correlations for stars with lower activity values but this seems not to be the case because we found significant correlations for the cases of \ion{Na}{i} and \ion{He}{i} for low values of $S$ (middle and right panels of Figure \ref{r_s_ha}). We therefore attribute this trend to the decrease of the importance of filaments and the increase of the contribution of plages to the H$\alpha$ index  as the activity level of the stars increases.

We found no dependence of the correlation between these indices and color. We note, however, that this was observed with very low signal-to-noise ratio for the (bluer) spectral regions that contain the \ion{Ca}{ii} H and K lines.

\subsection{$S_{\mathrm{Ca\,II}}$ vs $\ion{Na}{i}$}
\begin{figure*}[!Htb]
\begin{center}
\includegraphics[width=6cm]{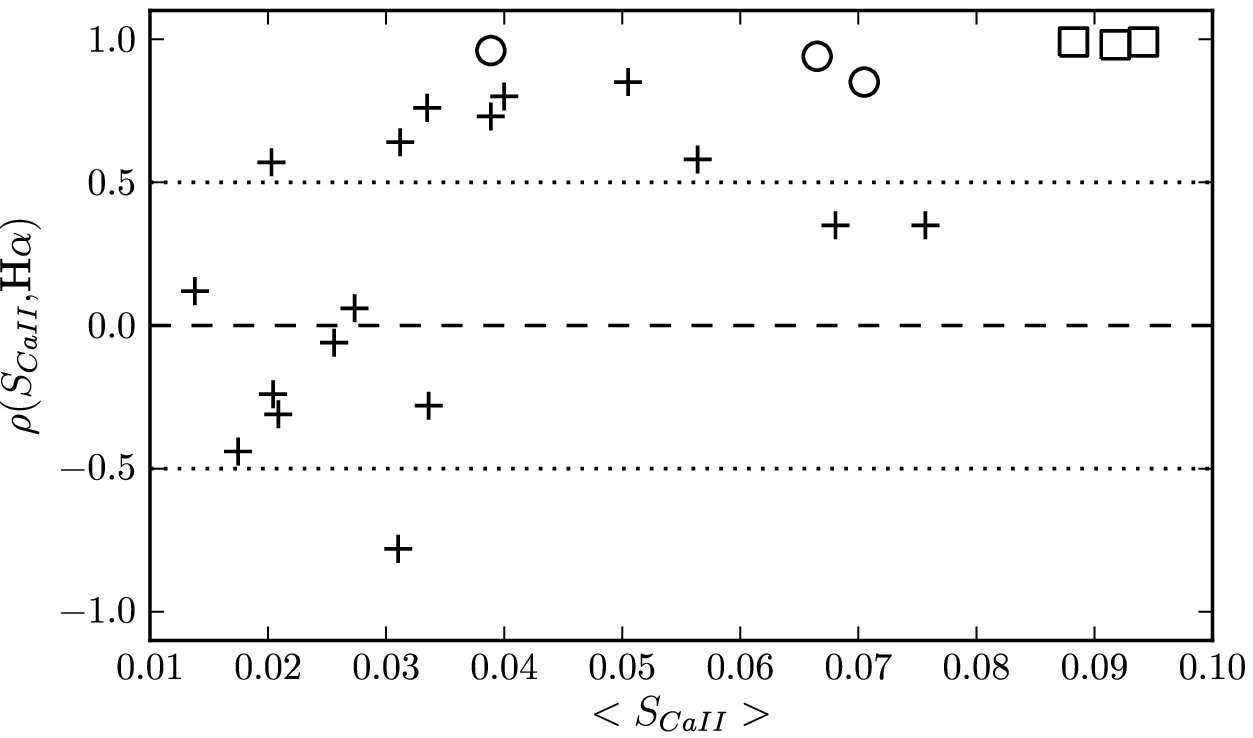}
\includegraphics[width=6cm]{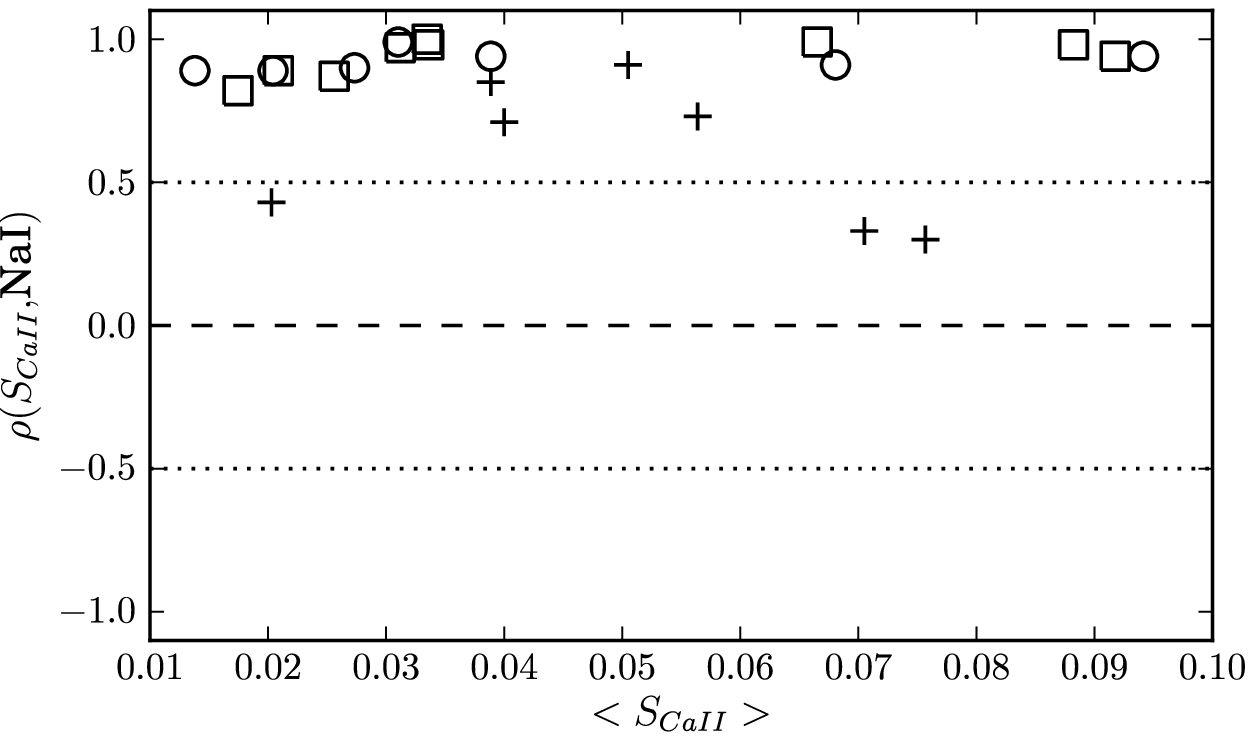}
\includegraphics[width=6cm]{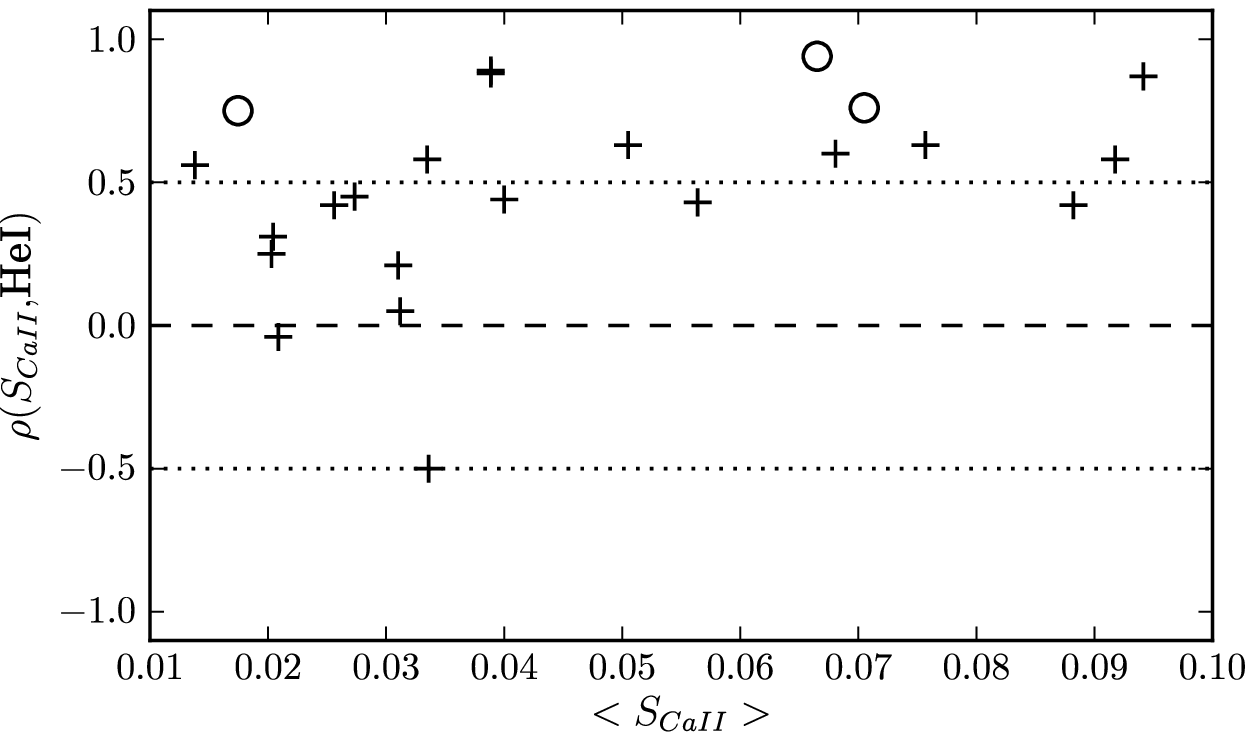}
\caption{Pearson correlation coefficients for the relations between $S_{\mathrm{Ca\,II}}$, H$\alpha$, \ion{Na}{i}, and \ion{He}{i} as a function of mean $S_{\mathrm{Ca\,II}}$  for the 23 stars with values of $S$ index. Open squares are values with FAP $\leq 0.01$, open circles for $0.01<$ FAP $\leq 0.05$, and "plus" ("$+$") symbols for FAP $> 0.05$.}
\label{r_s_ha}
\end{center}
\end{figure*}

Table \ref{table:ind}, Figure \ref{r_s_ha} (middle panel), and the time-series presented in Figures \ref{ind_bjd1}-\ref{ind_bjd6}, show that the \ion{Na}{i} behaves similarly to the $S_{\mathrm{Ca\,II}}$ index.

\citet{diaz2007a} studied the \ion{Na}{i} D1 and D2 lines for a sample of late-F to mid-M and found that these lines could be used as chromospheric activity indicators for very active stars. In our sample of low-activity stars, 16 out of 23  (around 70\% of the sample) have a strong correlation (with FAP $\leq$ 0.05) between the $\ion{Na}{i}$ index and our $S$-index. We have no negative values of the correlation coefficient and all the stars except three have $\rho > 0.5$. Note that neither of these three stars passed the variability test for these two indices. Furthermore, we found no relation between the correlation coefficient and the average values of $S$, $\ion{Na}{i}$, or with $V-I$. This means that the $\ion{Na}{i}$ lines can be used as a proxy of the long-term activity as measured by $\ion{Ca}{ii}$ for low-activity early-M dwarfs.
This can be important because the \ion{Ca}{ii} H \& K lines are located in the blue part of the spectrum and in these late-type stars the flux in this region is very low compared to the redder regions where the $\ion{Na}{i}$ D1 and D2 lines are located.
That \citet{diaz2007a} have not found this trend for the least active stars of their sample might be because they used medium-resolution spectra, while we used spectra obtained with high resolution.

\subsection{$S_{\mathrm{Ca\,II}}$ vs $\ion{He}{i}$}
When we studied the variability of the activity indices in Section \ref{var}, we noted that the $\ion{He}{i}$ index was not showing as much variability as the other three indices. Only three stars present a variability with P-value below 0.05. Now, when we studied the correlation coefficient between this index and the $S_{\mathrm{Ca\,II}}$, we observed that the correlation between the two is very weak. Only three stars show correlation below a FAP value of 0.05 and the correlation coefficients vary between $-0.50$ and 0.94. This further confirms that the $\ion{He}{i}$ index is a poor activity proxy (as measured by the $S$-index) for early-M dwarfs. Figure \ref{r_s_ha} (right panel) shows the variation of the coefficient with mean $S$-index. Although there is a tendency for positive correlations as shown in the other indices, the coefficients are more scattered than for the $S$--$\ion{Na}{i}$ case and less significant. A similar trend as the one observed for H$\alpha$ can also be observed here, where for stars with lower $<S_{\mathrm{Ca\,II}}>$ there is a tendency for lower or negative values of the coefficient.

\subsection{$\mathrm{H\alpha}$ vs $\ion{Na}{i}$, $\mathrm{H\alpha}$ vs $\ion{He}{i}$, and $\ion{Na}{i}$ vs $\ion{He}{i}$}
Because the correlation between the calcium and the sodium lines is very strong for almost all stars in our sample, it is expected that the relation between these two indices and H$\alpha$ will be similar.
This is the case, because the majority of the stars that show strong correlations here also have strong correlations between $S$ and the hydrogen line, except for Gl581, which shows an anti-correlation for the two cases.
Therefore, the discussion in Section \ref{s_ha} could also make sense in this case.

Nine stars out of 30 show strong a correlation between H$\alpha$ and $\ion{He}{i}$ (around 30\% of the sample) with four more stars having $0.05 <$ FAP $\leq 0.1$.
This shows that the $\ion{He}{i}$ index is more similar in behavior to H$\alpha$ than to $S_{\mathrm{Ca\,II}}$, where only three out of 23 stars showed a strong correlation.

As expected from the $S$--$\ion{He}{i}$ comparison, the correlation between the $\ion{Na}{i}$ and $\ion{He}{i}$ indices is weak. Only 17\% of the stars have FAP values lower than 0.05.

\subsection{Nightly averaged correlations}
Although the aim of this work was not to study the correlations between the four indices on short timespans (our concern was with long-term, cycle-type variations), we also calculated the correlation coefficients for the relations $S_{\mathrm{Ca\,II}}$--H$\alpha$ and $S_{\mathrm{Ca\,II}}$--\ion{Na}{i} for the nightly averaged data. The correlations observed for the binned data were maintained. The trend observed for the correlation between $S$ and H$\alpha$ when plotted against average $S$ is also present and \ion{Na}{i} maintains a very good correlation with the $S$ index that is independent of the activity level. The FAP values in general were lower because here we had more data points and therefore it was more difficult to obtain higher correlations by permutating individual points. Again, this result supports the use of the \ion{Na}{i} index as a proxy of activity of M dwarfs even for high-frequency variations as does the fact that the H$\alpha$ activity indicator does not vary linearly with activity as measured by the $S$ index.

This study will be presented in more detail in a future paper.

\begin{figure*}[!Htb]
\begin{center}
\includegraphics[width=6cm]{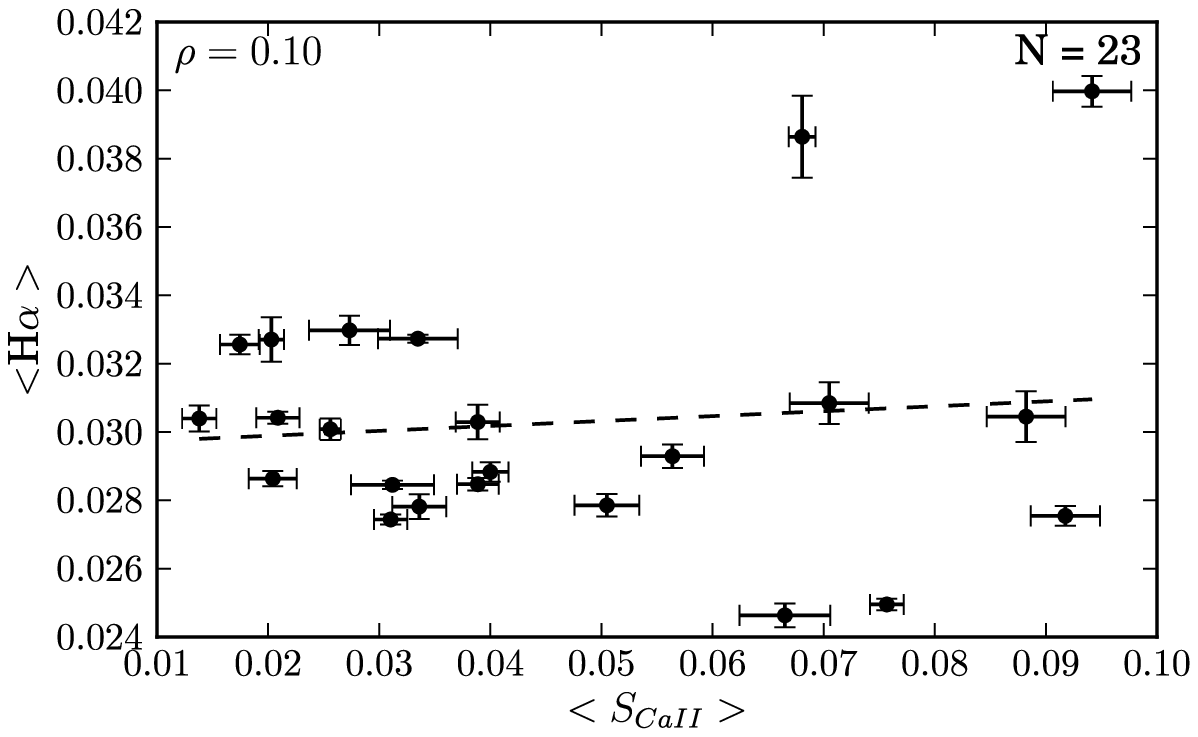}
\includegraphics[width=6cm]{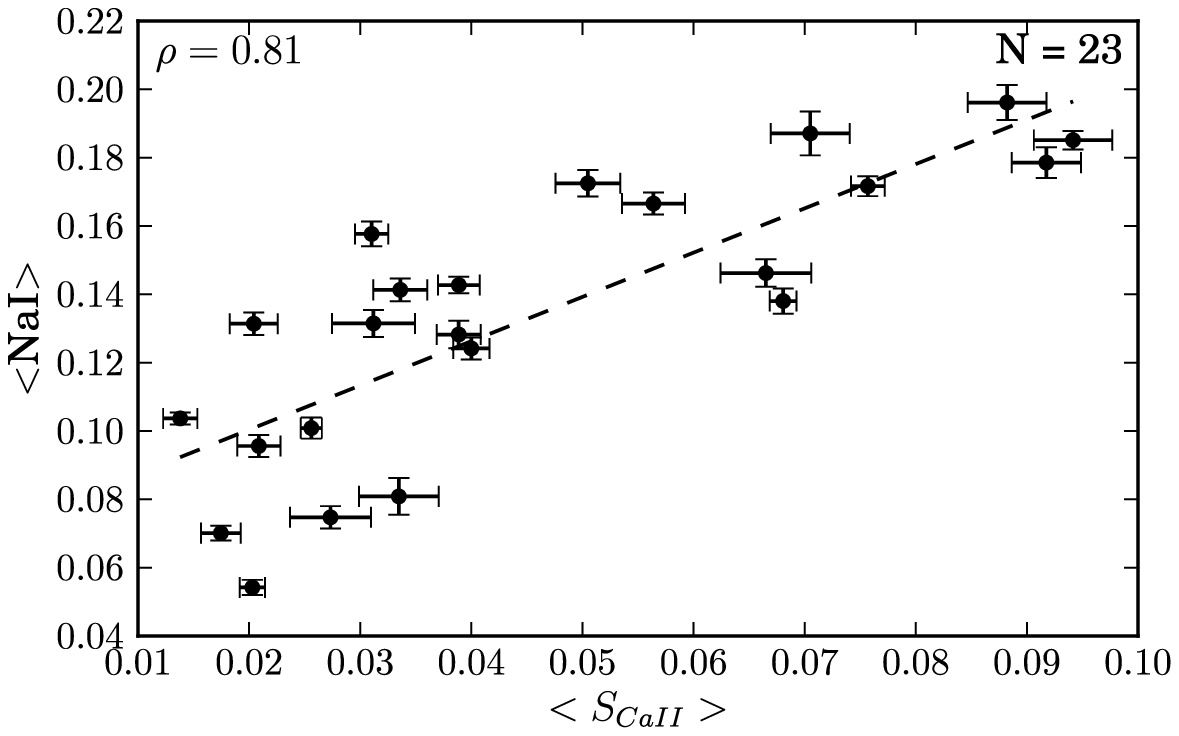}
\includegraphics[width=6cm]{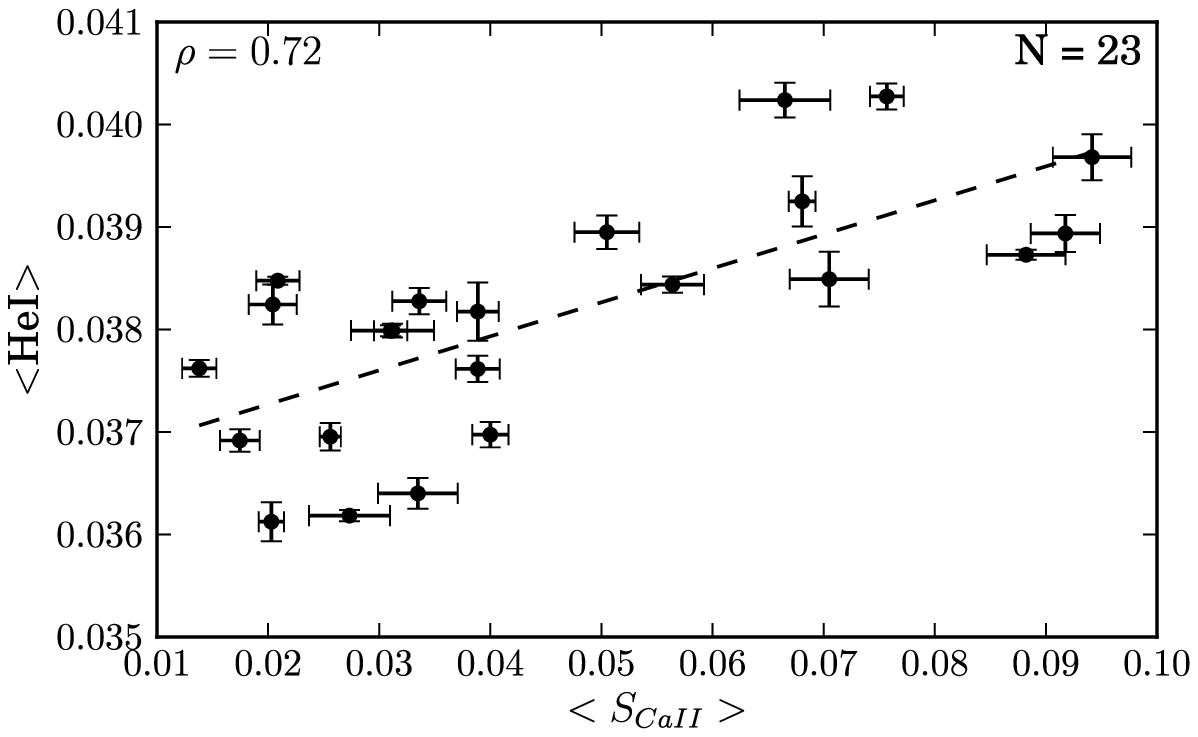}
\caption{Comparison between the average values of $S_{\mathrm{Ca\,II}}$, $\ion{H\alpha}{}$, $\ion{Na}{i}$, and $\ion{He}{i}$ for the 23 stars with values of $S$ index. Errorbars are the standard deviations, the dashed line the best linear fit, $\rho$ is the Pearson correlation coefficient, and $N$ the number of stars used.}
\label{mean_ind}
\end{center}
\end{figure*}

\begin{figure*}[!Htb]
\begin{center}
\includegraphics[width=6cm]{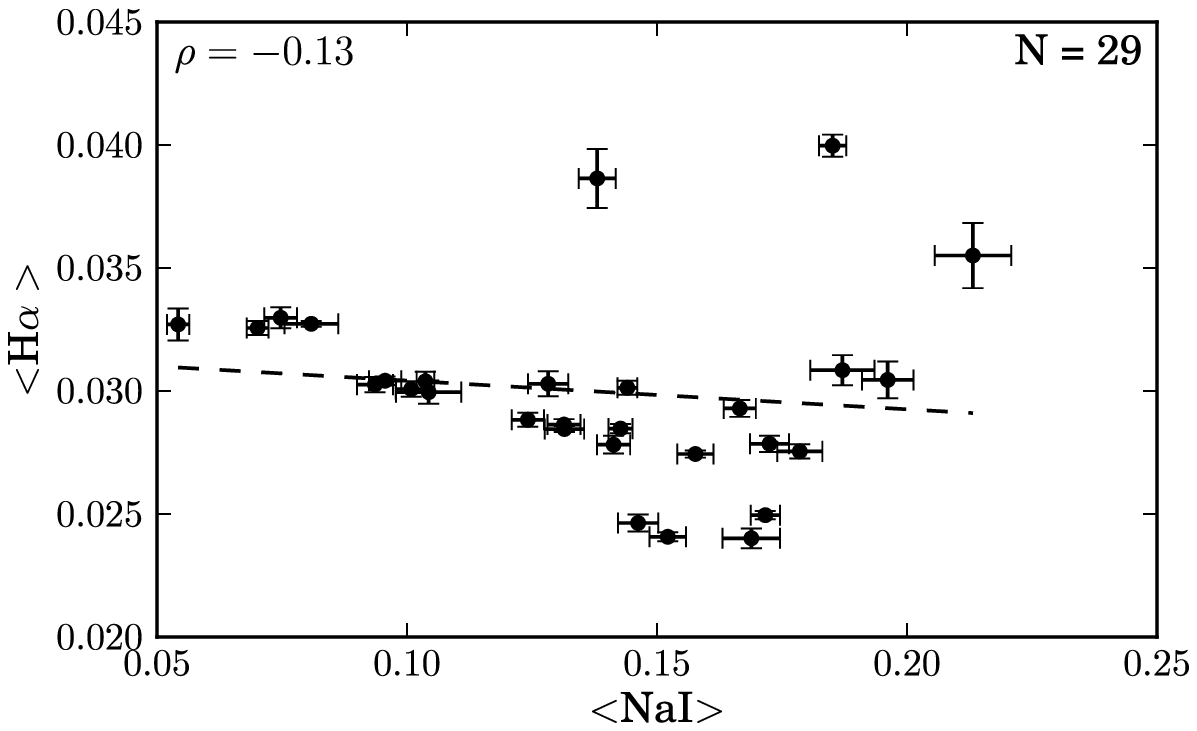}
\includegraphics[width=6cm]{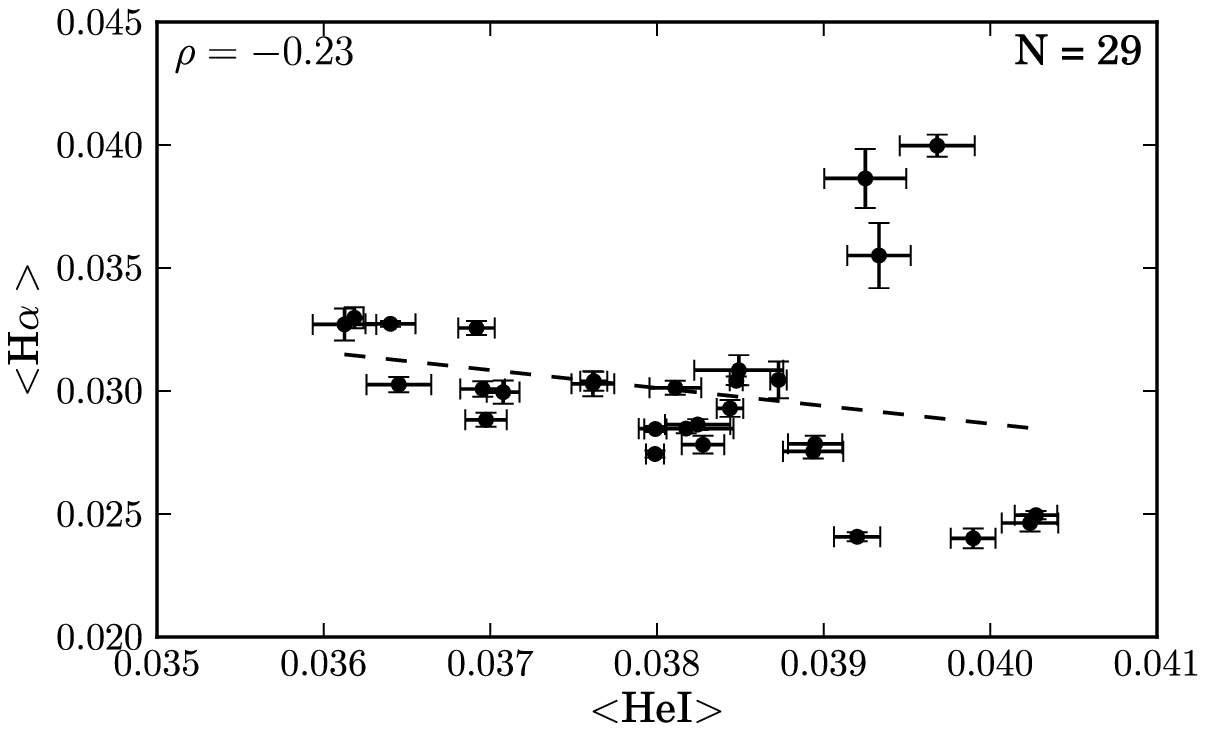}
\includegraphics[width=6cm]{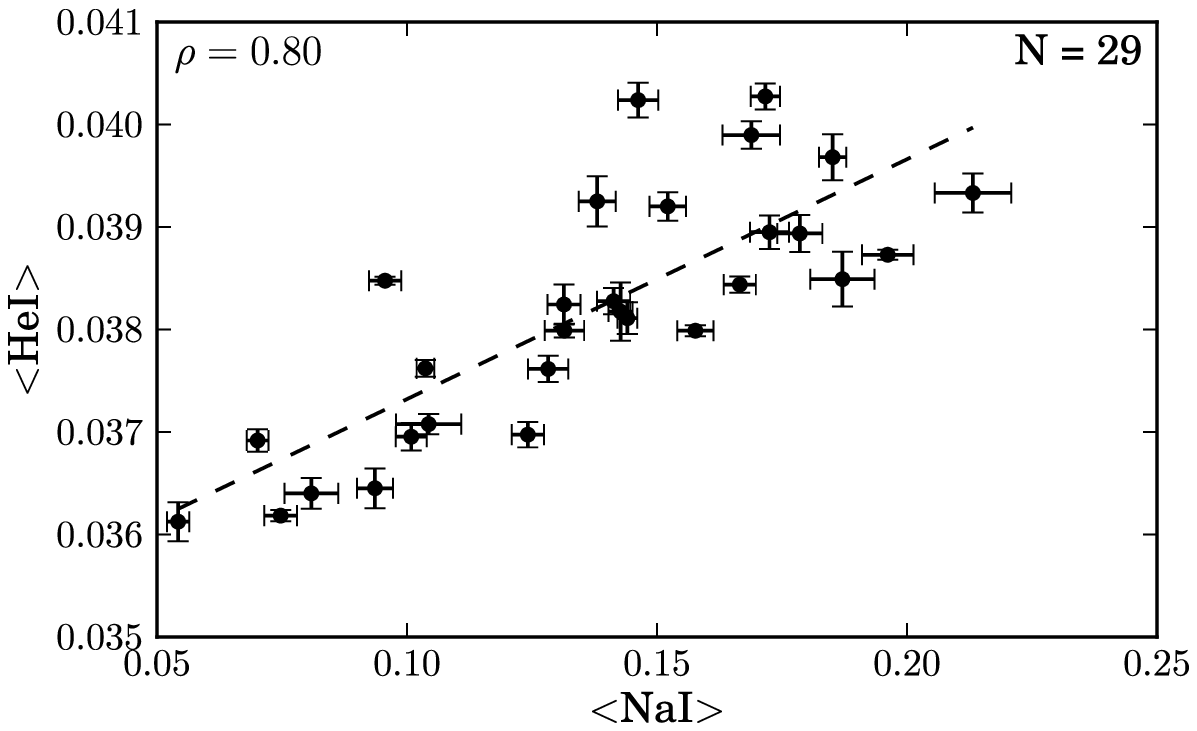}
\caption{Comparison between the average values of $\ion{H\alpha}{}$, $\ion{Na}{i}$, and $\ion{He}{i}$ for 29 stars. Gl551 was excluded from these plots because it has very high activity values. Symbols as in Figure \ref{mean_ind}.}
\label{mean_ind2}
\end{center}
\end{figure*}

\subsection{Mean activity indices and color} \label{ind_col}
\begin{figure}[!Htb]
\begin{center}
\includegraphics[width=4.4cm]{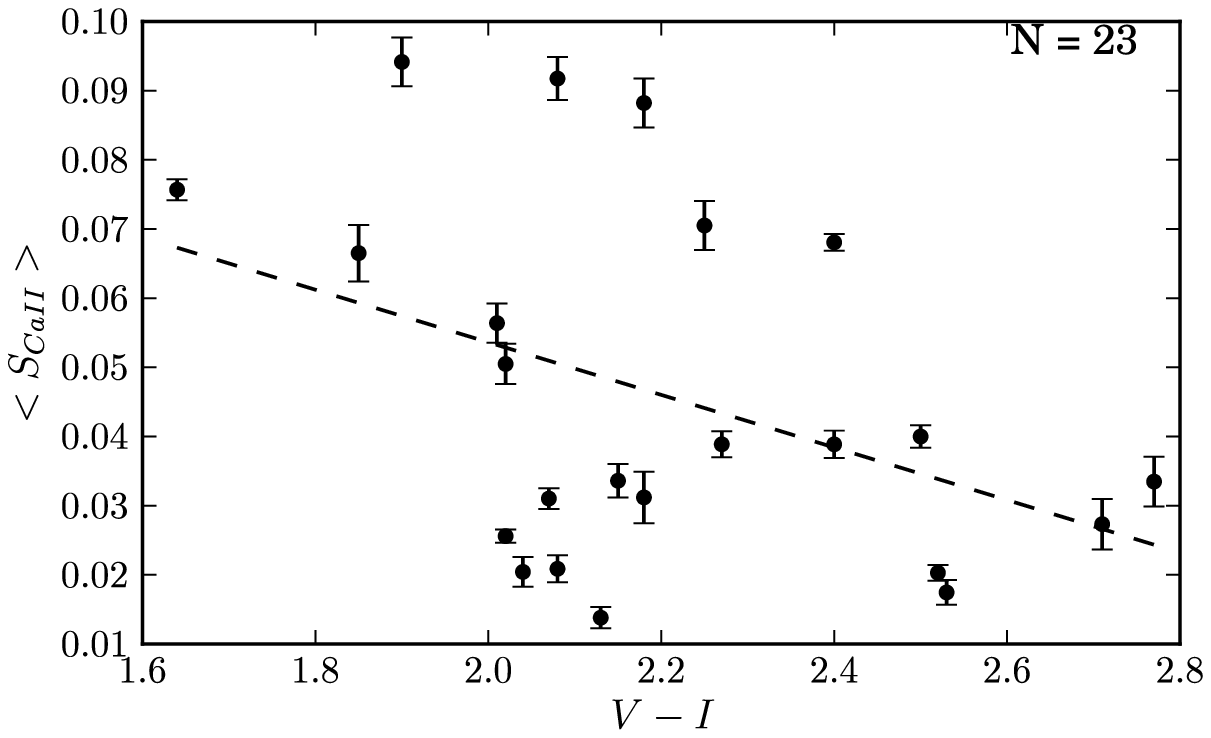}
\includegraphics[width=4.4cm]{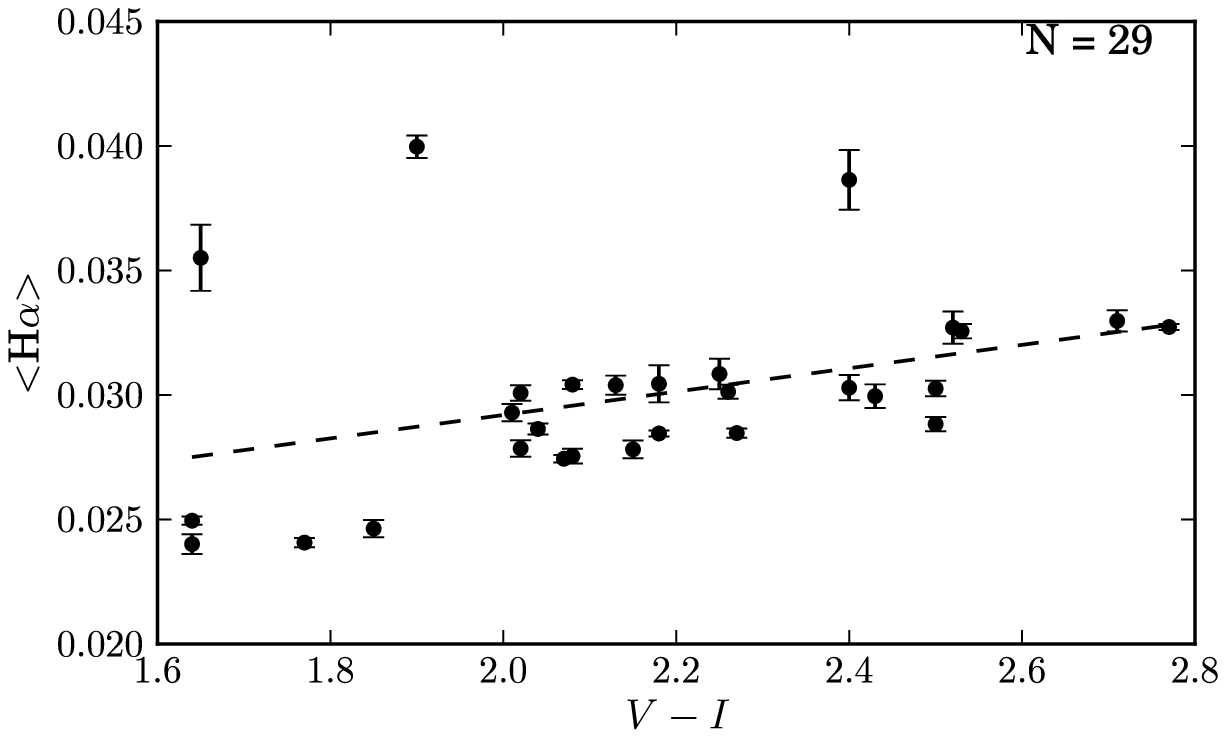}
\includegraphics[width=4.4cm]{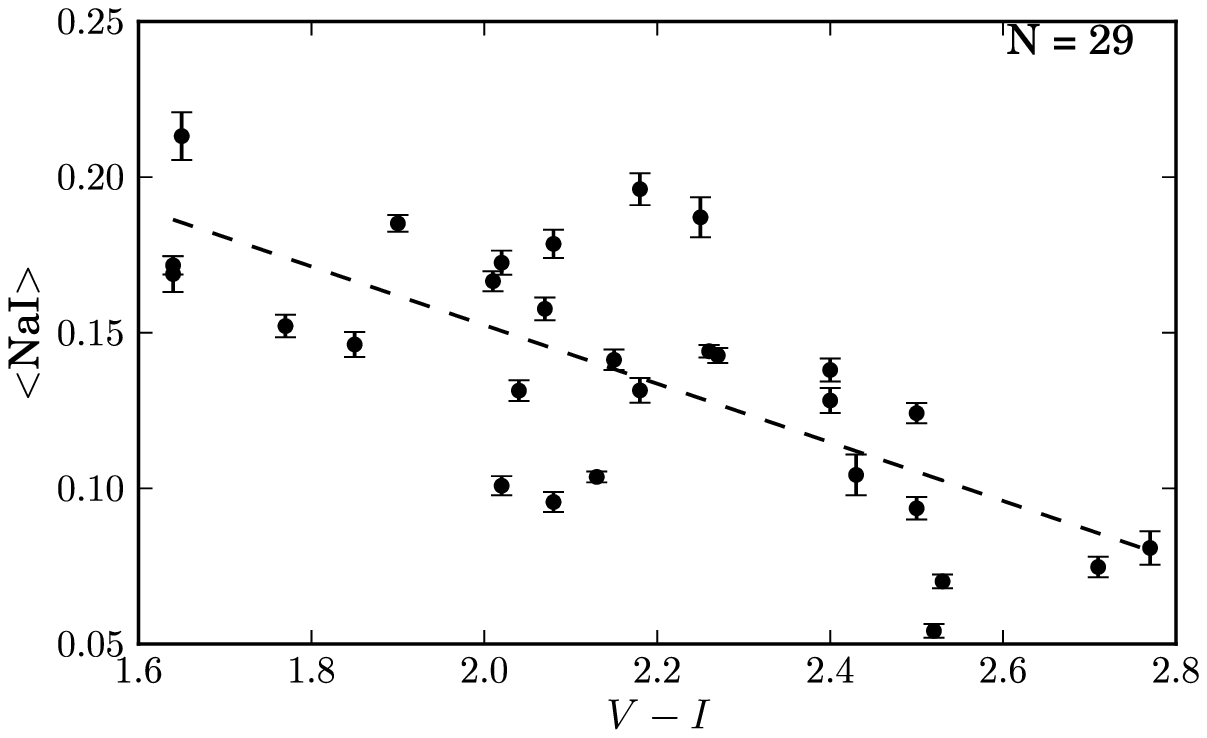}
\includegraphics[width=4.4cm]{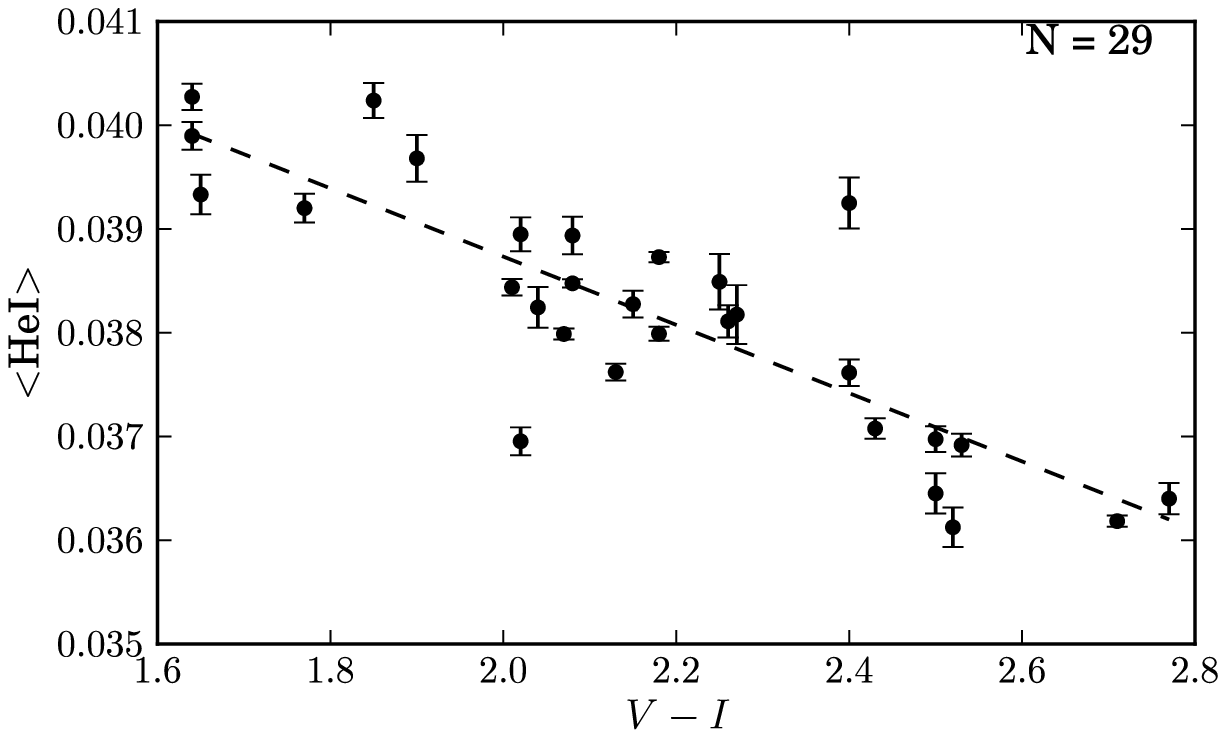}
\caption{Relation between the four activity indices and $V-I$ color. Symbols as in Figure \ref{mean_ind}.}
\label{colour}
\end{center}
\end{figure}

We compared the average values of the four activity indicators of each star to look for correlations (Figures \ref{mean_ind} and \ref{mean_ind2}).
We found that the mean values of $S_{\mathrm{Ca\,II}}$, $\ion{Na}{i}$ and $\ion{He}{i}$ are linearly correlated, but the same is not true when we compared them with the hydrogen line.
Because these activity proxies were not corrected for the contribution of photospheric flux, we expect these indices to be dependent on the spectral type when we compare stars with different colors.
As we can see in Figure \ref{colour}, the same three indices that have a linear correlation between their average values appear to have a similar trend with $V-I$: all of them decrease with color.
On the other hand, the H$\alpha$ index increases with $V-I$.

\begin{figure*}[!htb]
\begin{center}
\includegraphics[width=6cm]{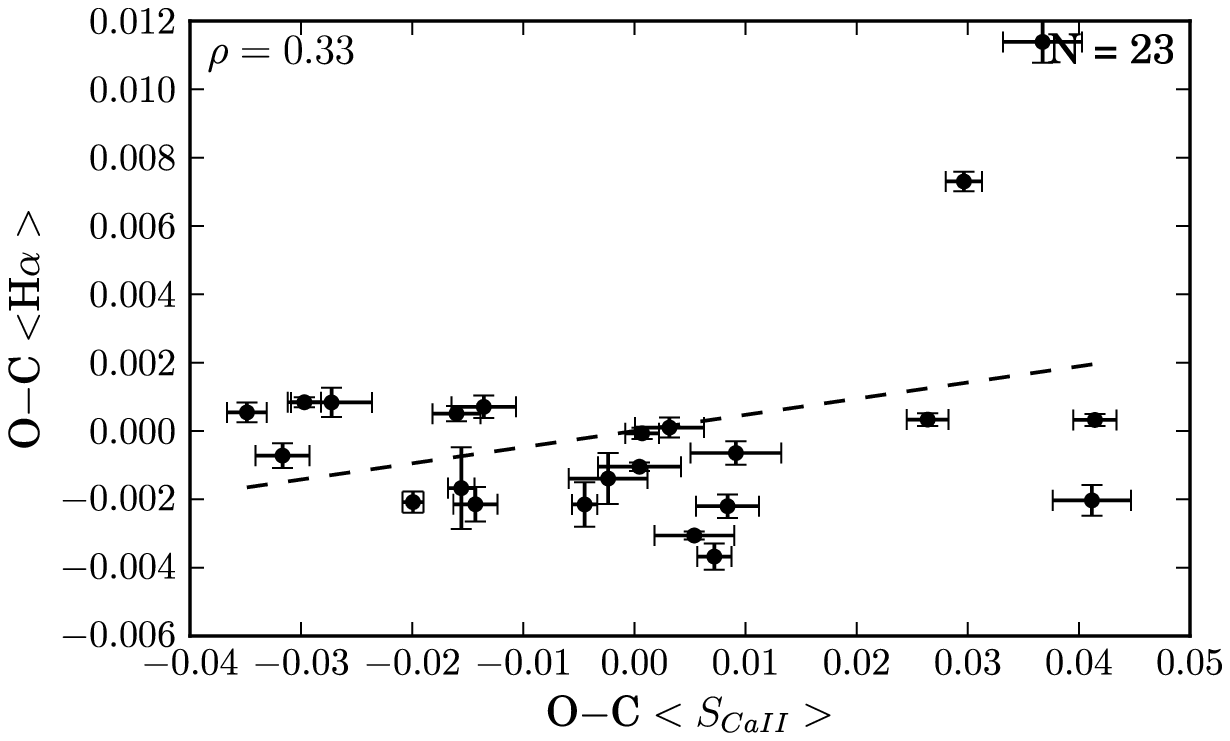}
\includegraphics[width=6cm]{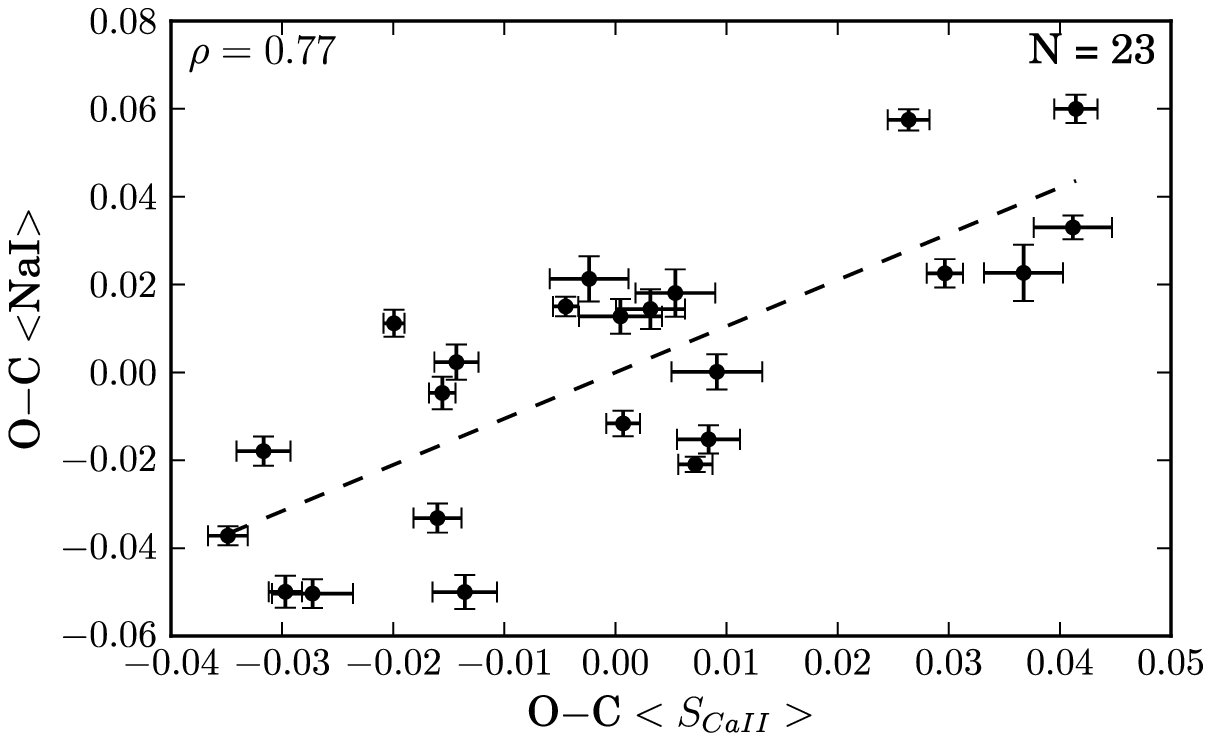}
\includegraphics[width=6cm]{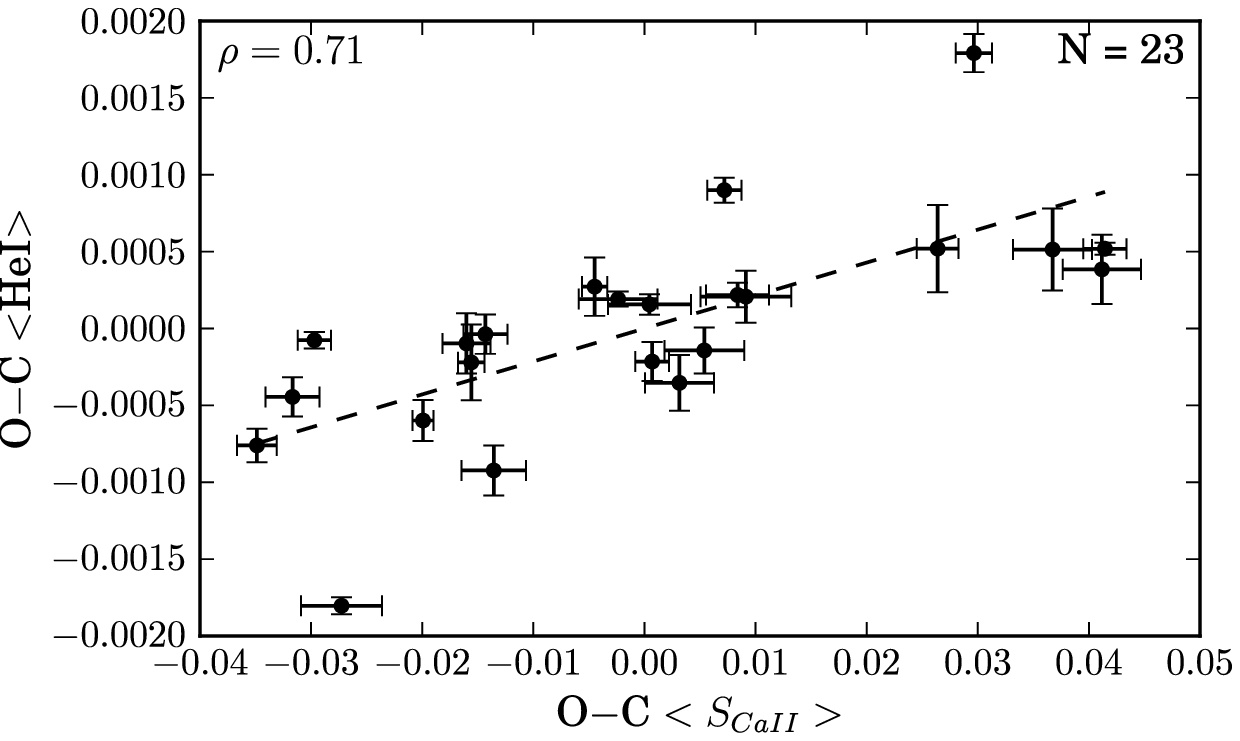}
\caption{Comparison between the average values of $S_{\mathrm{Ca\,II}}$, $\ion{H\alpha}{}$, $\ion{Na}{i}$, and $\ion{He}{i}$ for the 23 stars with values of $S$ index after correction for color. Symbols as in Figure \ref{mean_ind}.}
\label{mean_ind_corr}
\end{center}
\end{figure*}

\begin{figure*}[!htb]
\begin{center}
\includegraphics[width=6cm]{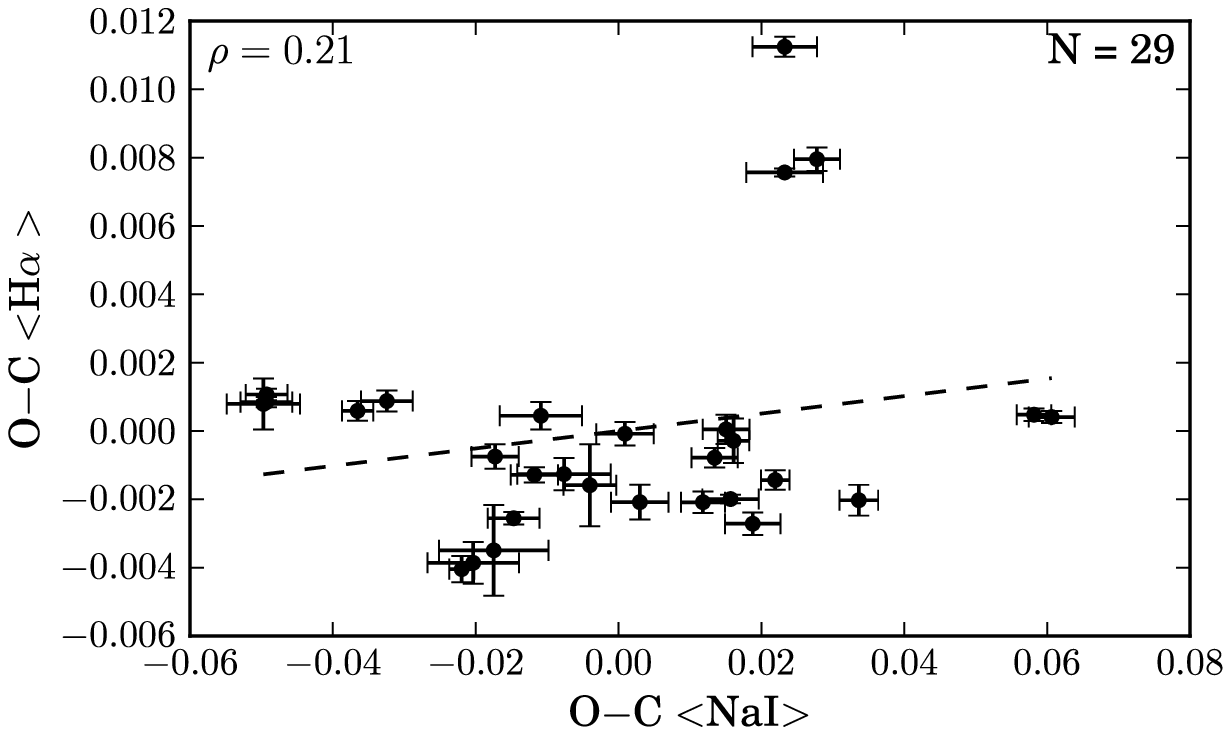}
\includegraphics[width=6cm]{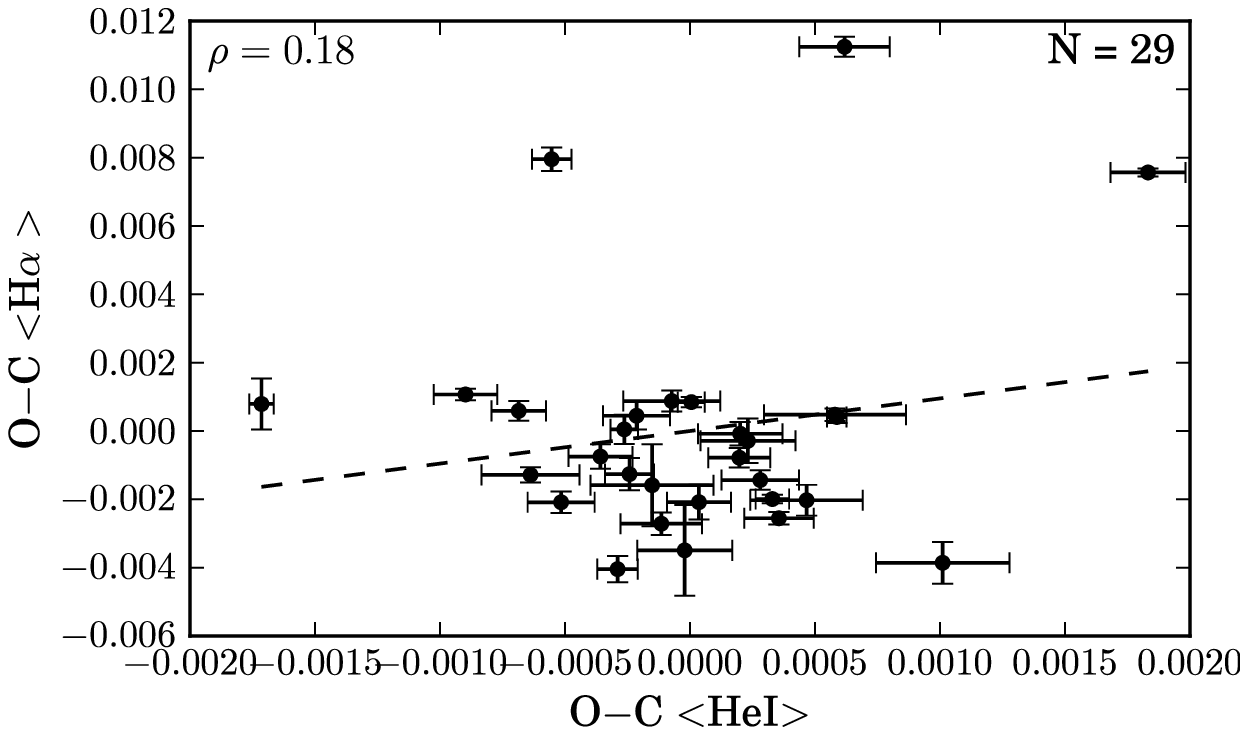}
\includegraphics[width=6cm]{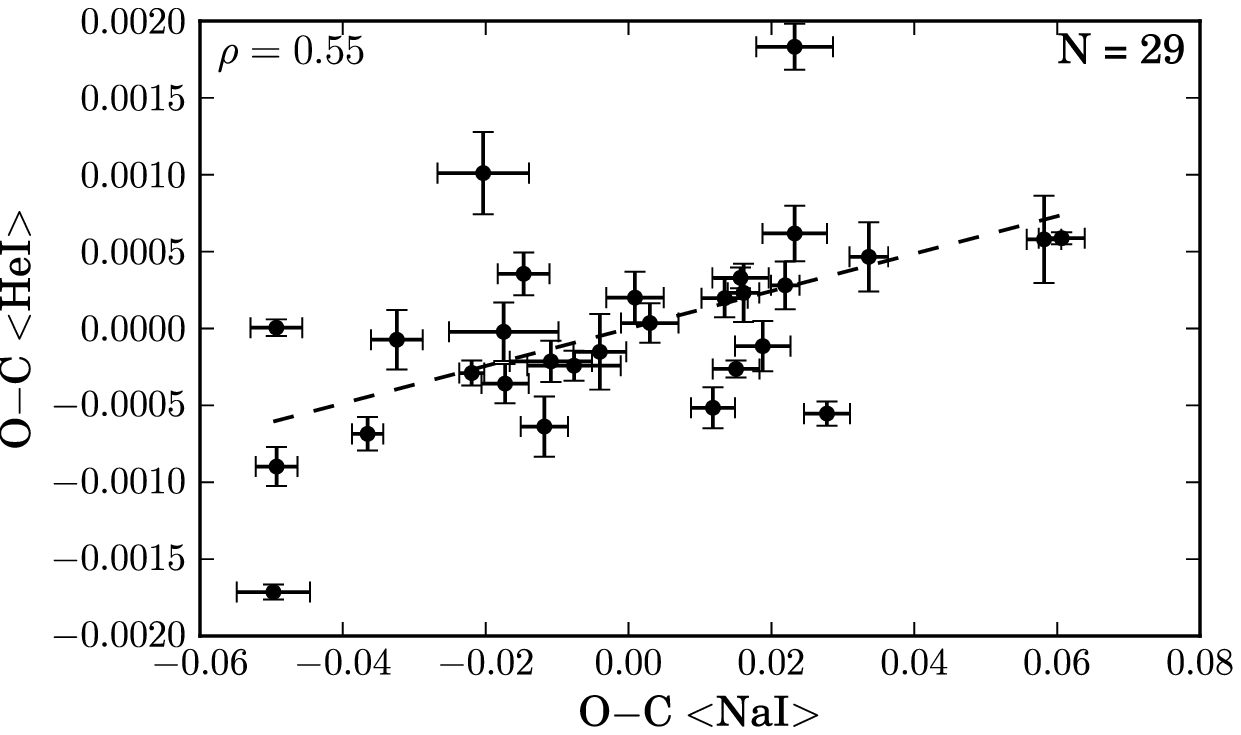}
\caption{Comparison between the average values of $\ion{H\alpha}{}$, $\ion{Na}{i}$, and $\ion{He}{i}$ for 29 stars after correction for color. Gl551 was excluded from these plots because of its very high activity values. Symbols as in Figure \ref{mean_ind}.}
\label{mean_ind_corr2}
\end{center}
\end{figure*}

We corrected the average values of the four indices using the residuals of the best-fit lines shown in Figure \ref{colour}.
Figures \ref{mean_ind_corr} and \ref{mean_ind_corr2} show the relations between the residuals of the indices after correction.
The linear trends between $<\ion{Na}{i}>$ vs $<S>$ and $<\ion{He}{i}>$ vs $<S>$ were maintained with similar correlation coefficient values except for $<\ion{He}{i}>$ vs $<\ion{Na}{i}>$, which decreased $\rho$ from 0.80 to 0.55.

Although the average values of these three indices are correlated, the scatter is still high.
It is known that the fraction of active M dwarfs is not constant with spectral type but, increases for spectral types later than M4 \citep{delfosse1998}. This could be a source of biasing toward the redder end of spectral color. However, in this sample only two stars have spectral type later than or equal to M4 and the redder of them (Gl551, with spectral type M5.5) was discarded from these corrections.

It is clear from Figures \ref{mean_ind_corr} and \ref{mean_ind_corr2} that direct comparisons between H$\alpha$ and the other three indices are not possible because there is no correlation between them.
This reinforces the idea that the hydrogen line does not measure activity in the same way as $\ion{Ca}{ii}$, $\ion{Na}{i}$, or $\ion{He}{i}$ for early-M dwarfs.
The strong intra-star correlations observed between the $S$ and $\ion{Na}{i}$ lines are also observed when the activity level of different stars is compared. Therefore, the sodium lines can be used not only to measure the relative activity variations for a given star, but also to compare activity between stars.

\section{Conclusions}\label{conclusion}
We used more than five years of HARPS high-resolution spectra of a sample of M0--M5.5 stars from the HARPS M-dwarf planet search survey to study the behavior of the $S_{\mathrm{Ca\,II}}$, H$\alpha$, $\ion{Na}{i}$, and $\ion{He}{i}$ chromospheric activity indicators.
The data were binned to average-out unwanted short time-scale variations. We first addressed the question of the mid- to long-term variability of the activity indices using F-tests. We found that some stars in our sample were showing obvious signs of these variations in $S_{\mathrm{Ca\,II}}$, H$\alpha$, and \ion{Na}{i}. The same was not as evident for the \ion{He}{i} index. The most obvious cases of detected long-term activity variations are Gl1, Gl273, Gl433, Gl436, Gl581, Gl588, Gl667C, Gl832, Gl849, Gl877, Gl908, and HIP85647. This selection was made by choosing stars with statistically significant variation in at least two indices.

We also performed F-tests to detect possible maxima or minima of activity. Gl1 and Gl433 have statistically significant maxima in at least two indices. Gl581 appears to be in a statistically significant minimum of activity in all indices except for H$\alpha$, which shows a maximum.

Our next step was to compare the activity indicators using the Pearson correlation coefficient. Similarly to what was found by \citet{cincunegui2007b} we obtained a great variety of correlations between the $S_{\mathrm{Ca\,II}}$ and H$\alpha$ in the range $-0.8 < \rho < 1$. We found that there is a trend of the correlation coefficient with the level of activity as measured by the $S$-index. For the least active stars there is a tendency for low- or anti-correlation, while for the most active stars in our sample the correlation becomes strongly positive.

We attribute this effect to the fact that as activity increases, the H$\alpha$ lines first start as a weak absorbing line, then become strongly absorbing, and as the activity further increases, they start to fill in, becoming weak absorbers again, and finally at high-activity levels become a full emission line, as predicted by \citet{cram1979} (see also \cite{cram1987} \cite{stauffer1986}, \cite{walkowicz2009}). A relation like this would produce the correlation observed in Figure \ref{r_s_ha} (left panel). This ambiguous relation between the calcium and H$\alpha$ lines was also observed in Figure \ref{lines} where we compared the shape of the activity lines for different activity levels.
This behavior might be a consequence of the different contribution of filaments to both indices, their different spatial distribution in the stellar disk, and the fact that its contribution to the H$\alpha$ index saturates at a given level of activity \citep[see][]{meunier2009}.
We thus conclude that, although the H$\alpha$ is able to detect long-term activity variations in early-M dwarfs, it is not the same physical phenomena that is being probed by both indices and thus their variations cannot be naively compared. Nevertheless, the simultaneous study of $S_{\mathrm{Ca\,II}}$ (or \ion{Na}{i}) and H$\alpha$ might be useful to gather new information about the presence of different activity-related features such as filaments and plages in stellar disks.

The \ion{He}{i} index has not only the smallest variation of the three lines, we also found it to correlate less well with activity and thus we do not recommend its use for the long-term activity study of M dwarfs.

We found a very good correlation between $S_{\mathrm{Ca\,II}}$ and \ion{Na}{i}. A great number of stars display a similar behavior for both indices and there is no relation between the correlation coefficient and the activity level. This index consequently is a proxy of activity variations as measured by $S_{\mathrm{Ca\,II}}$. Because these stars have more flux in the redder part of the spectrum, the \ion{Na}{i} index should be preferred over the $\ion{Ca}{ii}$ lines because the measurements will have a far better signal-to-noise ratio. We can then extend the \citet{diaz2007a} suggestion of the use of these lines to measure mid- to long-term activity in the least active stars of later spectral type.

In a following paper we will compare the behavior of the long-term activity from these results with the radial-velocity measurements and the cross-correlation function parameters bisector span, full-width-at-half-maximum, and contrast, with the aim of investigating if long-term cycle-like activity variations are able to influence the measured radial velocity of M dwarfs.

\begin{acknowledgements}
We would like to thank our anonymous referee for the helpful comments and suggestions. This work has been supported by the European Research Council/European Community under the FP7 through a Starting Grant, as well as in the form of a grant reference PTDT/CTE-AST/098528/2008, funded by Funda\c{c}\~ao para a Ci\^encia e a Tecnologia (FCT), Portugal. J.G.S. would like to thank the financial support given by FCT in the form of a scholarship, namely SFRH/BD/64722/2009. N.C.S. would further like to thank the support from FCT through a Ci\^encia 2007 contract funded by FCT/MCTES (Portugal) and POPH/FSE (EC).
\end{acknowledgements}

\bibliographystyle{aa} 
\bibliography{magcyclesM_ind} 

\end{document}